\documentclass{article}
\usepackage[utf8]{inputenc}
\usepackage[english]{babel}
\usepackage{amsmath}
\usepackage{amssymb}
\usepackage{algorithm}
\usepackage{algorithmic}
\usepackage{verbatim}
\usepackage{enumitem}
\usepackage{multirow}
\usepackage{graphicx,color}
\usepackage[outercaption]{sidecap}
\usepackage[numbers,sort]{natbib}
\usepackage{etoolbox}
\usepackage[pdftitle={Fast event-based simulation},
  pdfauthor={Bauer/Engblom/Widgren},
  pdffitwindow=true,
  breaklinks=true,
  colorlinks=true,
  urlcolor=blue,
  linkcolor=red,
  citecolor=red,
  anchorcolor=red]{hyperref}

\newcommand{\X}{\mathbb{X}}

\newcommand{\task}{\mathcal{T}}

\newcommand{\rand}{\mbox{rand}}

\newcommand{\Nspecies}{N_{c}}
\newcommand{\Ndet}{N_{d}}
\newcommand{\Nreactions}{N_{t}}
\newcommand{\Ncells}{N_{n}}

\newcommand{\Stoich}{\mathbb{S}}
\newcommand{\Ext}{\mathbb{R}}
\newcommand{\Connect}{\mathbb{C}}

\newcommand{\Stoichglobal}{\mathbb{G}}

\newcommand{\Realdom}{\mathbf{R}}
\newcommand{\Intdom}{\mathbf{Z}}
\newcommand{\Probspace}{\Omega}
\newcommand{\Probelem}{\omega}
\newcommand{\Probfiltr}{\mathcal{F}}
\newcommand{\Prob}{\mathbf{P}}

\newcommand{\fatmu}{\boldsymbol{\mu}}
\newcommand{\fatnu}{\boldsymbol{\nu}}
\newcommand{\fatlambda}{\boldsymbol{\Lambda}}

\newcommand{\review}[1]{\textcolor{black}{#1}} 
\newcommand{\secreview}[1]{\textcolor{black}{#1}} 

\numberwithin{equation}{section}
\numberwithin{table}{section}
\numberwithin{figure}{section}

\title{Fast event-based epidemiological simulations on
  national scales}

\author{Pavol Bauer$^{\mbox{\tiny{1}}}$ \and Stefan
  Engblom$^{\mbox{\tiny{1}}}$ \thanks{Corresponding author:
    S. Engblom, telephone +46-18-471 27 54, fax +46-18-51 19 25.} \and
  Stefan Widgren$^{\mbox{\tiny{2}}}$ \\
  [2ex] \parbox[c]{5cm}{\centering
    \footnotesize{\textit{$^{\mbox{\tiny{\rm{1}}}}$Division of
        Scientific Computing \\ Department of Information Technology
        \\ Uppsala University \\ SE-751 05 Uppsala, Sweden. \\ email:
        \texttt{pavol.bauer, stefane@it.uu.se,} \\
  }}} \parbox[c]{5cm}{\centering
    \footnotesize{\textit{$^{\mbox{\tiny{\rm{2}}}}$Department of
        Disease Control and Epidemiology \\ National Veterinary
        Institute \\ SE-751 89 Uppsala, Sweden. \\ email:
        \texttt{stefan.widgren@sva.se}}}}}

\date{January 27, 2016}

\begin{document}

\maketitle

\begin{abstract}
  We present a computational modeling framework for data-driven
  simulations and analysis of infectious disease spread in large
  populations. For the purpose of efficient simulations, we devise a
  parallel solution algorithm targeting multi-socket shared memory
  architectures. The model integrates infectious dynamics as
  continuous-time Markov chains and available data such as animal
  movements or aging are incorporated as externally defined events.

  To bring out parallelism and accelerate the computations, we
  decompose the spatial domain and optimize cross-boundary
  communication using dependency-aware task scheduling. Using
  registered livestock data at a high spatio-temporal resolution, we
  demonstrate that our approach not only is resilient to varying model
  configurations, but also scales on all physical cores at realistic
  work loads. Finally, we show that these very features enable the
  solution of inverse problems on national scales.

\vspace{0.25cm}


\vspace{0.25cm}


%

\end{abstract}


\section{Introduction}

\review{Livestock diseases have a major economic impact on farmers, the
livestock industry and countries \cite{hasonova2006,
  Knight-Jones2013}. Modeling and simulation of infectious disease
spread is important in designing cost-efficient surveillance and
control \cite{Willeberg2011}. One challenge is that disease dynamics
and transmission routes for various pathogens are fundamentally
different. Indirect transmission of pathogens via the environment for
fecal-oral diseases requires a different model compared to diseases
that spread with direct contact between individuals
\cite{BrooksPollock2014}. Another challenge is to incorporate the
increasing amount of epidemiologically relevant data into the models
\cite{Pellis2015}. It is therefore desirable to have simulation tools
that are flexible to various disease spread models yet efficient to
handle the large amounts of available livestock data.}

\review{Due to uncertainties in the exact details in pathogen
  transmission \cite{greenwood_stochastic_2009} and the inherent
  random nature of animal interactions, stochastic modeling is natural
  and often required. Spatial models that include proximity to
  infected farms with local clustering of disease spread gained
  popularity during the Foot-mouth-disease epidemic in 2001
  \cite{keeling_models_2005,
    stevenson_interspread_2013,harvey_north_2007}. Another important
  route for disease spread is animal trade, creating a temporal
  network of contacts between farms \cite{masuda_temporal_2013}. It
  has been shown that the topology and connectivity of the network has
  great impact on the disease spread and on the effect of control
  measures \cite{shirley_topology_2005, buettner_trade_2014}.}

Stochastic models on discrete state-spaces are typically simulated
using Discrete Event simulation (DES), a general approach to evolve
dynamical systems consisting of discrete events including, in
particular, continuous-time Markov chains (CTMCs)
\cite{cassandras_systems_2008}. As most realistic epidemiological
models are formulated on a large state-space and/or need to be studied
over comparably long periods of time, parallelization is
desirable. The highest degree of parallelism is typically achieved by
a decomposition of the spatial information, often represented as a
graph or network, into a set of sub-domains
\cite{fujimoto_parallel_1990}. It is then up to the strategy for event
handling at domain boundaries how well the concurrent execution scales
and which overall degree of parallelism is extractable. As it may
hinder scalability, a constraint that plays a crucial role in the
design of parallel DES is to maintain the sequential ordering of
events, that is, to preserve the underlying causality of the model.

In general, there are two types of boundary events that can occur
during a simulation, which hence ultimately decide what will be the
optimal parallelization strategy. Those which are deterministic and
essentially of fully predictable character, and those which are
stochastic and not predictable at an earlier simulation time
\cite{fujimoto_parallel_1999}. To parallelize events that belong to
the latter group, sophisticated approaches such as optimistic parallel
DES algorithms have been proposed \cite{jefferson_virtual_1985}. These
approaches may use speculative execution to enable scalability but
must implement rollback mechanisms in case the event causality is
violated \cite{carothers_efficient_1999}. Alternatively, in
simulations where the domain crossing events are deterministic and
thus predictable, conservative simulation may be used as it is
possible to avoid causal violations \review{altogether
  \cite{fujimoto_parallel_1990, heidelberger_conservative_1993}}. In
particular, a parallel scheduler \cite{drozdowski_parallel_2009} can
be used to create an execution order which guarantees causality,
\review{as has been previously shown in
  \cite{xiao_scheduling_1999,nicol_composite_2002}, notably with the
  focus on simulation of telecommunication networks.}

\review{In this paper we present an efficient and flexible framework
  for data-driven modeling of disease spread simulations. The model
  integrates disease dynamics as continuous-time Markov chains and
  real livestock data as deterministic events. This allows us to
  create a temporal network of disease transmission, which has been
  shown to be a key aspect in modeling and simulation of spatial
  disease spread \cite{shirley_topology_2005, buettner_trade_2014}.
  Previously, agent-based simulations based on synthetic data have
  been studied by others
  \cite{barrett_episimdemics_2008,yeom_overcoming_2014}.}

\review{The way the model is defined allows us to predict future
  boundary events at any simulation time, and hence we are able to
  create parallel execution traces which respect causality.  In
  particular, we find that dependency-aware task computing
  \cite{subhlok_exploiting_1993,leijen_design_2009} can be used to
  implement this approach with high efficiency, as all the necessary
  information to maintain spatial and temporal causality of events can
  be specified via dynamic creation of tasks and dependencies.}

\review{This is in contrast to previous approaches
  \cite{xiao_scheduling_1999,nicol_composite_2002}, as the scheduler
  is not an implicit part of the parallel simulation algorithm, but
  can be chosen by the user from a wide selection of openly available
  libraries (e.g.~\texttt{Open MP} 4.0, \texttt{OmpsS}
  \cite{duran_ompss:_2011}, or \texttt{StarPU}
  \cite{augonnet_starpu:_2011}). We show how the selected library is
  integrated into our simulation framework, by assigning parts of the
  sequential algorithm to independent tasks that are scheduled using a
  certain set of rules. We evaluate this approach using the
  task-parallel run-time library \texttt{SuperGlue}
  \cite{tillenius_superglue_2014}, which has been demonstrated to be
  an efficient scheduler of fine-grained tasks.}  Using our simulator
on models with realistic work-loads, we demonstrate scalability on a
multi-socket shared-memory system and investigate when this approach
is preferable in comparison to traditional parallelization
techniques. As the achievable scalability clearly depends on the
properties of the individual model, we in particular choose to
investigate the influence of the model's connectivity pattern.

The paper is organized as follows. In \S\ref{sec:model} we introduce
the mathematical foundation for our framework. In
\S\ref{sec:implementation} we discuss the sequential simulation
algorithm and the strategy for parallelization. In
\S\ref{sec:experiments} we present numerical experiments carried out
on benchmarks consisting of a recently proposed epidemiological model
incorporating large amounts of registered data. We also include an
example of an inverse problem for an epidemic model on national
scales. Finally, in \S\ref{sec:conclusion} we offer a concluding
discussion around the central themes of the paper.


\section{Epidemiological modeling}
\label{sec:model}

We consider in this section a highly general approach to
epidemiological modeling. Proceeding stepwise we start with a
description of single-node stochastic SIR-type models in the form of
continuous-time Markov chains, using a compact notation that also
encompasses externally defined events. We next couple an ensemble of
such single-node models into a network with prescribed transitions in
between the nodes to arrive at a global description. Finally, since
most realistic models on multiple scales will typically incorporate
also quantities for which a continuous description is more natural, we
consider a \emph{mixed} approach in which continuous-time Markov
chains are coupled to ordinary differential equations (ODEs).

\subsection{Discrete states}

We shall use a compact notation for jump stochastic differential
equations (jump SDEs) as follows. We assume a probability space
$(\Probspace,\Probfiltr,\Prob)$ where the filtration
$\Probfiltr_{t \ge 0}$ contains Poisson processes of any finite
dimensionality. The time dependent \emph{state vector}
$X_{t} = X(t; \, \omega) \in \Intdom_{+}^{\Nspecies}$, with
$\Probelem \in \Probspace$, counts at time $t$ the number of
individuals of each of $\Nspecies$ different categories, or
\emph{compartments}. Since the random process is of discrete
character, the map $t \to X(t)$ is right continuous only; by $X(t-)$
we therefore denote the value of the state \emph{before} any events
scheduled at time $t$.

Given a \emph{rate function} $r: \Intdom_{+}^{\Nspecies} \to
\Realdom_{+}$ and a \emph{stoichiometric coefficient} $s \in
\Intdom^{\Nspecies}$, we write a continuous-time Markov chain in the
form
\begin{align}
  \label{eq:scalarJSDE}
  dX_{t} &= s\mu(dt),
\end{align}
with scalar counting measure $\mu(dt) = \mu(r(X(t-)); \, dt)$. This
notation expresses a dynamics consisting of events with exponentially
distributed waiting times of intensities $r(X(t-))$; specifically
$E[\mu(dt)] = E[r(X(t-)) \, dt]$. An event at time $t$ implies that
the state is to be changed according to the prescription $X(t) =
X(t-)+s$. In \eqref{eq:scalarJSDE}, note that if some stoichiometric
coefficient $s_{i} < 0$, then we must have that $r(x) = 0$ for $x_{i}$
small enough, or otherwise the chain will reach negative states with
positive probability.

The generalization of \eqref{eq:scalarJSDE} to non-scalar counting
measures is straightforward. Assuming $\Nreactions$ different
transitions specified by a vector intensity $R:
\Intdom_{+}^{\Nspecies} \to \Realdom_{+}^{\Nreactions}$ and a
stoichiometric matrix $\Stoich \in \Intdom^{\Nspecies \times
  \Nreactions}$, we simply write
\begin{align}
  \label{eq:vectorJSDE}
  dX_{t} &= \Stoich\fatmu(dt),
\end{align}
with $\fatmu(dt) = [\mu_{1}(dt),\ldots,\mu_{\Nreactions}(dt)]^{T}$
and, for each $k$, $\mu_{k}(dt) = \mu(R_{k}(X(t-)); \, dt)$.

As a concrete example, consider the classical SIR-model
\cite{Kermack1927}
\begin{align}
\label{eq:sirtrans}
  &\left. \begin{array}{rl}
    S+I &\xrightarrow{\beta} 2I \\
    I &\xrightarrow{\gamma} R \\
  \end{array} \right\}
  \intertext{With state vector $X = [S,I,R]$ this can be understood as}
  \label{eq:SIRstoich}
    \Stoich &= \left[ \begin{array}{rr}
        -1 & 0 \\
        1  & -1 \\
        0 & 1
      \end{array} \right], \\
    R(x) &= [\beta x_{1}x_{2},\gamma x_{3}]^{T}.
\end{align}

With one small additional convention the above notation also
encompasses events that have been defined externally. Suppose, for
example, in the SIR-model, that susceptible individuals are to be
added one by one at known deterministic times $(t_{i})$. To accomplish
this we replace \eqref{eq:SIRstoich} with
\begin{align}
 \label{eq:sirtrans_ext}
  \Stoich &= \left[ \begin{array}{rrr}
      -1 & 0 & 1 \\
      1  & -1 & 0 \\
      0 & 1 & 0
    \end{array} \right], \\
  \intertext{and additionally define in terms of the Dirac measure,}
  \mu_{3}(dt) &= \sum_{i} \delta(t_{i};\,dt).
\end{align}
Eq.~\eqref{eq:vectorJSDE} now evolves the full dynamics of the coupled
stochastic-deterministic model. Note that when \emph{removing}
individuals using this scheme, some care is required to be able to
guarantee a non-negative chain.


\subsection{Network model}
\label{sec:network}

Although the previous discussion is of completely general character it
makes sense to handle the collective dynamics of a possibly very large
collection of nodes in a slightly more streamlined fashion. Assuming
$\Ncells$ nodes in total we consider the state matrix $\X \in
\Intdom_{+}^{\Nspecies \times \Ncells}$ and evolve the local dynamics
by a version of \eqref{eq:vectorJSDE},
\begin{align}
  \label{eq:local}
  d\X^{(i)}_{t} &= \Stoich\fatmu^{(i)}(dt).
\end{align}

Given an undirected graph $\mathcal{G}$ each node $i$ is modeled to
affect the state of the nodes in the connected components $C(i)$ of
$i$, and in turn, to be affected by all nodes $j$ such that $i \in
C(j)$. The interconnecting dynamics can then be written as
\begin{align}
  \label{eq:global}
  d\X^{(i)}_{t} &= -\sum_{j \in C(i)} \Connect\fatnu^{(i,j)}(dt)+
  \sum_{j; \, i \in C(j)} \Connect\fatnu^{(j,i)}(dt).
\end{align}
Note that in \eqref{eq:global}, global consistency is enforced as
follows. The $k$th ``outgoing'' event is a change of state according
to $\X^{(i)}(t) = \X^{(i)}(t-)-\Connect_{k}$, and, for some $j \in
C(i)$, $\X^{(j)}(t) = \X^{(j)}(t-)+\Connect_{k}$. By inspection the
intensity for this transition is $E[\nu^{(i,j)}_{k}(dt)] =
E[N_{k}^{(i,j)}(\X^{(i)}(t-)) \, dt]$, say, where the dependency is
only on the state of the ``sending'' node $i$.

Using superposition of \eqref{eq:local} and \eqref{eq:global} the
overall dynamics becomes
\begin{align}
  \label{eq:master}
  d\X^{(i)}_{t} &= \Stoich\fatmu^{(i)}(dt)-
  \sum_{j \in C(i)} \Connect\fatnu^{(i,j)}(dt)+
  \sum_{j; \, i \in C(j)} \Connect\fatnu^{(j,i)}(dt).
\end{align}
As before we conveniently allow externally defined deterministic
events to be included in this description using the equivalent
construction in terms of Dirac measures.

\subsection{Continuous states}

In the previous description we assumed essentially that
\emph{individuals} were counted, such that a discrete stochastic model
was needed to accurately capture the dynamics of a possibly small and
noisy population. In a multiscale model, however, it makes sense to
allow also \emph{continuous} state variables, representing, for
example, environmental properties more naturally described in a
macroscopic language.

Assuming an additional continuous state matrix $Y \in
\Realdom^{\Ndet \times \Ncells}$ to be available we find that a
general model corresponding to \eqref{eq:master} is
\begin{align}
  \label{eq:master_det}
  \frac{dY^{(i)}(t)}{dt} &= f(\X^{(i)}(t-),Y^{(i)}(t)) \\
  \nonumber &-
  \sum_{j \in C(i)} g(\X^{(i)}(t-),Y^{(i)}(t))+
  \sum_{j; \, i \in C(j)} g(\X^{(j)}(t-),Y^{(j)}(t)).
\end{align}
Importantly, with this addition \eqref{eq:master} can now depend on
the continuous state variable,
\begin{align}
  \label{eq:intens1}
  E[\mu_{k}^{(i)}(dt)] &= E[R_{k}(\X^{(i)}(t-),Y^{(i)}(t)) \, dt], \\
  \label{eq:intens2}
  E[\nu^{(i,j)}_{k}(dt)] &= E[N_{k}^{(i,j)}(\X^{(i)}(t-),Y^{(i)}(t)) \, dt],
\end{align}
where of course $k$ is in the range where the dynamics is stochastic
rather than defined externally from a database.

Eqs.~\eqref{eq:master}, \eqref{eq:master_det}, and
\eqref{eq:intens1}--\eqref{eq:intens2} form the basis for our
epidemiological computational framework, next to be described.


\section{Implementation}
\label{sec:implementation}

In the following section we discuss the implementation details of our
computational framework. We begin with indicating how numerical
methods can be consistently designed to approximate the mathematical
model arrived at previously. A description of the sequential solution
algorithm and a presentation of the chosen parallelization strategy
based on domain decomposition then follows. We propose to process
events that cross domain boundaries as tasks and thus conclude with
the introduction of dependency aware task computing and an associated
scheduling scheme.

\subsection{Numerical Methods}
\label{sec:numerics}

\review{In order to be able to effectively incorporate finitely
  resolved temporal data as well as to obtain a parallelizable
  framework, we discretize time as $0 = t_{0} < t_{1} < t_2 <
  \cdots$.} We thus write the epidemiological model in
\eqref{eq:master}--\eqref{eq:master_det} in integral form, using a
global notation which incorporates the whole network,
\begin{align}
  \label{eq:numerical}
  \X_{n+1} &= \X_{n} + \int^{t_{n+1}}_{t_n} \Stoichglobal \fatlambda (ds), \\
  \label{eq:numerical2}
  Y_{n+1} &= Y_{n} + \int^{t_{n+1}}_{t_n} F(\X(s),Y(s)) \, ds,
\end{align}
with the understanding that $(\X,Y)_{n} = (\X,Y)(t_{n})$.

Typical numerical approaches to
\eqref{eq:numerical}--\eqref{eq:numerical2} are constructed via
operator splitting and finite differences \cite{jsdesplit}. As a
representable example we take
\begin{align}
  \label{eq:numstep1}
  \X_{n+1} &= \X_{n} + \int^{t_{n+1}}_{t_n} \Stoichglobal
  \fatlambda(\X(s-),Y_{n}; \; ds), \\
  \label{eq:numstep2}
  Y_{n+1} &= Y_{n} + \int^{t_{n+1}}_{t_n} F\left([\X_{n}+\X_{n+1}]/2,
  Y(s) \right) \, ds.
\end{align}
In \eqref{eq:numstep1} we freeze the variable $Y$ at a previous
time-step and integrate the stochastic dynamics only. Next, in
\eqref{eq:numstep2} we insert an average effective value of $\X$ and
integrate the deterministic part using any suitable deterministic
numerical method.

To describe a more concrete numerical method, some assumptions are in
order. Firstly, in \eqref{eq:master} we assume that \emph{events
  connecting two nodes have been externally defined}. In particular,
this assumption is satisfied for the important case of domesticated
herds of animals who move between nodes due to human interventions
only. Secondly, in \eqref{eq:master_det} we put $g = 0$ and thus
\emph{remove all direct influence between continuous variables in
  connected nodes}. This is reasonable for macroscopic variables that
are not easily transported, like bacterias in soil, but could of
course be violated for other media like groundwater or air.

For this scenario we can write down a concrete numerical method
per node $i$ as follows,
\begin{align}
  \label{eq:numstep3}
  \tilde{\X}_{n+1}^{(i)} &= \X_{n}^{(i)} + \int^{t_{n+1}}_{t_n} \Stoich
  \fatmu_{s}^{(i)}(\tilde{\X}^{(i)}(s-),Y_{n}^{(i)}; \; ds), \\
  \label{eq:numstep4}
  \X_{n+1}^{(i)} &= \tilde{\X}^{(i)}_{n+1}+\int^{t_{n+1}}_{t_n} \Stoich
  \fatmu_{d}^{(i)}(\X^{(i)}(s-),Y_{n}^{(i)}; \; ds) \\
  \nonumber
  &\phantom{= \tilde{\X}^{(i)}_{n+1}}-\int^{t_{n+1}}_{t_n} \sum_{j \in C(i)}
  \Connect\fatnu_{d}^{(i,j)}(\X^{(i)}(s-),Y_{n}^{(i)}; \; ds) \\
  \nonumber
  &\phantom{= \tilde{\X}^{(i)}_{n+1}}+\int^{t_{n+1}}_{t_n} \sum_{j; \, i \in C(j)}
  \Connect\fatnu_{d}^{(j,i)}(\X^{(i)}(s-),Y_{n}^{(i)}; \; ds), \\
  \label{eq:numstep5}
  Y_{n+1}^{(i)} &= Y_{n}^{(i)} + f(\tilde{\X}_{n+1}^{(i)},
  Y_{n}^{(i)}) \, \Delta t_{n}.
\end{align}
In \eqref{eq:numstep3} the stochastic part (subscript $s$) of the
measure is evolved in time to produce the temporary variable
$\tilde{\X}$. Next, \eqref{eq:numstep4} incorporates all externally
defined deterministic events (subscript $d$), both locally on the
node, and according to the connectivity of the network. Finally,
\eqref{eq:numstep4} is just the usual Euler forward method in time
with time-step $\Delta t_{n} = t_{n+1}-t_{n}$ evolving the continuous
state $Y$. The particular splitting method
\eqref{eq:numstep3}--\eqref{eq:numstep5} forms the basis for much of
the results reported in \S\ref{sec:experiments}.

\subsection{External events}
\label{sec:events}

Similar to the epidemiological events \eqref{eq:SIRstoich}, the
external events modify the discrete state according to a transition
vector \eqref{eq:sirtrans_ext}, but at a pre-defined time $t$.

We divide external events into two types; events of type $E_1$ operate
on the state of a single node, while events of type $E_2$ operate on
the states of two nodes.  It is meaningful to distinguish between
these types of events as they are processed differently by the
parallel algorithm discussed later.  They are defined by a set of
attributes
\begin{align}
	\label{eq:event}
    E_1=\{\Ext,t,n,i\}, \\
    E_2=\{\Ext,t,n,i, j\},
\end{align}
where $t$ is the time of the event, $\Ext$ the transition vector, $n$
the number of individuals affected and $i$ and $j$ the indices of the
affected nodes. This is a minimal set of attributes which can be
further extended for specific models.  As an example, within the
context of the SIR model \eqref{eq:sirtrans} we can define a
\emph{birth event} $\{\Ext,t,n, i\}$ of type $E_1$ with the transition
vector $\Ext=[1,0,0]^T$.

In the actual implementation, the transition vector is a column of the
stoichiometric matrix \eqref{eq:sirtrans_ext} that is indexed by the
event.  When the event is processed at time $t$, it changes the state
of node $i$ according to
\begin{align}
   \X^{(i)}_{t+1}=\X^{(i)}_{t}+\Ext n.
\end{align}
The overall spatial domain of the model can be understood as a graph
$G=(V,E)$. The edges $E$ result from events of type $E_2$ acting on
source and destination nodes $\X^{(i)}$ and $\X^{(j)}$.

\subsection{Sequential simulation algorithm}
\label{sec:sequential}

The sequential simulation algorithm is divided into three parts; the
processing of stochastic events \eqref{eq:numstep3}, hereafter referred
to as the \emph{stochastic step}, the processing of external events
\eqref{eq:numstep4}, or \emph{deterministic step}, and the
\emph{update of the continuous state variable} \eqref{eq:numstep5}.
These steps are processed repeatedly in the above-mentioned order until
the simulation reaches the end.

The stochastic step (Algorithm~\ref{alg:main}, p.~\pageref{alg:main})
is an adaptation of Gillespie's Direct Method
\cite{gillespie_exact_1977}.  The algorithm generates a trajectory
from a continuous-time Markov chain.  At first, the rates
$\omega_n^{(i)}$ for all stochastic events $n=1 \ldots N_{t}$ are
evaluated in all nodes $\X_i, i=1 \ldots \Ncells$.  Then, in each node
we sum up transition rates into $\lambda^{(i)}=\sum^n \omega_n^{(i)}$.
Next, the algorithm uses \emph{inverse transform sampling} to obtain
an exponentially distributed random variable representing the next
stochastic event time $\tau^{(i)}$ for each node $\X^{(i)}$,
\begin{align}
	\label{eq:sampling}
  \tau^{(i)} = -\log(\rand)/\lambda^{(i)}.
\end{align}
Here, $rand$ denotes a uniformly distributed random number in the
range $(0, 1)$. To obtain the index of the stochastic event that
occurred \emph{within} the node $\X^{(i)}$ we generate a new random
number $rand$ and find $n$ such that
\begin{align}
	\label{eq:sampling2}
\sum_{j=1}^{n-1} \omega_j(\X^{(i)}) < \lambda^{(i)} \, \rand
  \le \sum_{j=1}^n \omega_j(\X^{(i)}).
\end{align}
When $n$ is found, we compute the state update according to the
transition matrix \eqref{eq:SIRstoich}, setting
$\X^{(i)}_{t+\tau}=\X^{(i)}_t+\Stoich_n$ and simulation time
$t=t+\tau^{(i)}$.  Finally, to obtain the new next event time, the
rate $\omega_n^{(i)}$ of the event just occurred and all its dependent
events need to be re-computed as in \eqref{eq:sampling}.  For fast
execution, these dependencies are stored in a \emph{dependency graph}
that is traversed at this stage. The algorithm repeats until a defined
stopping time is reached, where the \emph{external events} will next
be processed.

The deterministic step works as a \textit{read and incorporate}
algorithm. It moves through the list of external events and processes
them at the defined event time.  In particular, if the event specifies
a single compartment where the transition occurs, it can be directly
applied to $\X^{(i)}$.

Finally, the continuous state variable is updated. As discussed in
\S\ref{sec:numerics}, in this step different numerical methods can be
applied. Note that the thus updated continuous state generally affects
the rate of stochastic events $\lambda^{(i)}$. Thus, before the
simulation proceeds with the next iteration of the stochastic step,
the event times need to be be rescaled \cite{gibson_efficient_2000}
using
\begin{align}
 \tau^{(i)}_{\mathrm{new}} &= t+\left(\tau^{(i)}_{\mathrm{old}}-t\right)
  \frac{\lambda^{(i)}_{\mathrm{old}}}{\lambda^{(i)}_{\mathrm{new}}}.
\end{align}
The implementation of the algorithm is written in \texttt{C}.  The
overall design is inspired and partly adapted from the Unstructured
Mesh Reaction-Diffusion Master Equation (\texttt{URDME}) framework
\cite{engblom_simulation_2009,drawert_urdme:_2012}.

\subsection{Parallel simulation algorithm}
\label{sec:parallel}

The parallelization starts with a decomposition of the spatial domain
of the model understood as a graph $G=(V,E)$. The target of this graph
partitioning problem is to divide the set of vertices $V$ of size
$\Ncells$ into $k$ approximately equally sized sub-domains
$V_1, V_2,..,V_k$. The cutting of edges $E$ follows straightforwardly
from the consecutive assignment of vertices to sub-domains. This
partitioning strategy does not guarantee a minimum amount of edge
cuts, but as the distribution of edges is predominantly homogeneous in
our data, we believe that the partitioning will not benefit from more
sophisticated approaches.  Nonetheless, if edges are distributed
heterogeneously, a Minimum Bisection algorithm
\cite{andreev_balanced_2004} may generate an optimized cut that
contributes to a better performance of the parallel solver.

After partitions $V_{1..k}$ are defined, the preprocessing algorithm
continues to rearrange the external events into a structure that is
more convenient for parallel processing.  Firstly, the external events
of type $E_1$ are assigned to $k$ lists $\mathcal{E}^k_1$, such that
all $E_1$ events affecting the nodes $\X^{(i)} \in V_k$ are stored in
the $k$th list.

Second, external events of type $E_2$ are divided in two categories;
external events of type $E_2$ where the source and destination nodes
lie within the same sub-domain $V_k$ are assigned to lists
$\mathcal{E}^k_2$. Events in lists $\mathcal{E}^k_1$ and
$\mathcal{E}^k_2$ can be processed by a thread assigned to the $k$th
sub-domain in private.  Events of type $E_2$ where the source node and
destination node do not lie within the same sub-domain $V_k$, are
assigned to a second list $\mathcal{E}^c_2$. This list then contains
\emph{domain crossing} external events that have to be handled by the
simulator in a special way.

The complexity of the data-rearrangement is $O(n)$, where $n$ is the
number of external events. Although $n$ can be large, the workload is
typically negligible for real-scale models. For example, in the
national scale model presented in \S\ref{sec:vtec}, the operation
takes $\sim0.1\%$ of the total simulation time on one core and
$\sim1\%$ of the simulation time on 32 cores, respectively.  Moreover,
re-arranged lists can be stored and re-used for models that are
simulated using the same decomposition, as is done in the simulation
study in \S\ref{sec:fitting}.

Finally, the decomposed problem can be simulated in parallel.  For
simplicity, let us assume that each sub-domain $k$ is bound to one
computing thread.  Then every thread processes the stochastic step
\eqref{eq:numstep3} and the update of the continuous variable
\eqref{eq:numstep5} on private nodes of $V_k$, as well as the
deterministic step \eqref{eq:numstep4} on external event lists
$\mathcal{E}^k_1$ and $\mathcal{E}^k_2$ (Algorithm~\ref{alg:parallel},
p.~\pageref{alg:parallel}). \review{Since time has been discretized,
  these computations are \emph{embarrassingly parallel}, in that no
  communication between neighboring threads is necessary during the
  processing of the $n$th time window $t_n+\Delta t_n$}. The potential
bottleneck of the simulation lies in the simulation of the
cross-boundary events in $\mathcal{E}^c_2$.

In our study we handle events in $\mathcal{E}^c_2$ in different ways.
The first possibility is to compute them entirely in serial.  This is
a valid approach if there are very few events in $\mathcal{E}^c_2$ in
relation to the private events as the scaling of the private
computations will not be affected.  On the other hand, if the overall
simulation is dominated by the processing of $\mathcal{E}^c_2$ events,
it can be regarded as serialized, as little concurrency will be
extractable using such an approach. Hence we focus on an intermediate
ratio of private and global work, where events in $\mathcal{E}^c_2$
occur at every deterministic time step $\Delta t$, but at a lower
frequency than the other private events.  We will also investigate if
scaling in this regime is achievable if $\mathcal{E}^c_2$ events are
scheduled using dependency-aware task-computing.

\subsection{Task-based computing}
\label{sec:taskbased}

An increasing amount of scientific computations are parallelized using
task-based computing
\cite{haidar_novel_2014,meng_scalable_2014,berry_event-based_2012}.
In order to apply this pattern the programmer typically has to divide
a larger chunk of work into a group of smaller tasks which can be
processed asynchronously. \review{A run-time library
  \cite{subhlok_exploiting_1993,leijen_design_2009}} is then used to
create an execution schedule of the tasks on the available parallel
hardware.

If the granularity of tasks is sufficiently fine, the schedule will be
denser and the idle time shorter. On the other hand, the scheduler
synchronizes a larger number of small tasks which usually implies more
overhead. \review{See \cite{gerasoulis_granularity_1993} for a
  thorough discussion on the impact of granularity}.

If the scheduler \review {supports \emph{dependency-awareness}
  \cite{perez_dependency-aware_2008}}, the programmer can further
define a number of task dependencies.  This is a critical feature if
data is shared between tasks and therefore a processing order has to
be enforced.  The scheduler then manages the dependencies in the form
of a Directed Acyclic Graph (DAG) and spawns tasks whenever all
dependencies are met.

We believe that the usage of task-based computing is beneficial in our
computational framework, as a small granularity of processes is given
by the underlying modeling. In our approach we aim to divide our
computations into tasks and define a scheduling policy which
guarantees causality of events although they are processed in
parallel.

These scheduling rules can be implemented on any dependency aware task
scheduler, \review{the only requirement for some of the scheduling
  policies is the support for dynamic addressing of a sub-set of
  dependencies, e.g. via an array of pointers. For example,
  \texttt{Open MP} 4.0 does not support this \cite{ompmanual}.}  In
our computational experiments, we make use of the run-time library
\texttt{SuperGlue} \cite{tillenius_superglue_2014}.  In
\texttt{SuperGlue}, dependencies are assigned to data and expressed
via \emph{data versioning } \cite{zafari_programming_2012}.  If a
chunk of data is being processed by a task, a version counter
representing the data access will be increased.  Other tasks that are
dependent on the chunk will be spawned whenever the new version
becomes available. \texttt{SuperGlue} has been demonstrated to be an
efficient shared-memory task-scheduler that it is capable of operating
at a comparably low synchronization overhead. The processing of
dependencies and spawning of tasks is dynamic, and \texttt{SuperGlue}
additionally supports load balancing by work stealing from
over-utilized threads.

\subsection{Scheduling and dependencies}
\label{sec:scheduling}

We now define the tasks and their dependencies that are used in the
task-based implementation of the parallel algorithm.  Task
$\task_{S}(k,n)$ executes private computations on the decomposed data
of the $k$th sub-domain (lines 5 and 6 of Pseudo-code
\ref{alg:parallel}). That is the stochastic step \eqref{eq:numstep3}
on all nodes $\X_n \in V_k$, the processing of the private external
events in lists $\mathcal{E}^k_1$ and $\mathcal{E}^k_2$
\eqref{eq:numstep4}, as well as the update of the continuous variables
\eqref{eq:numstep5}.  The counter $n$ indicates the iteration of the
time window $t_{n+1}=t_n+\Delta t_n$. 

Task $\task_{M}$ processes state updates due to the domain crossing
events stored in list $\mathcal{E}^c_2$.  In order to estimate the
possible impact of granularity of $\task_{M}$ tasks, we compare two
different scheduling policies; if the task is constructed for
\emph{coarse-grained processing}, we compute all $\mathcal{E}^c_2$
events occurring at the $n$th time window $\Delta t$ in one single
task.  Thus, the task takes only one argument, $\task_{M}(n)$.

If tasks are constructed for \emph{fine-grained processing}, we
schedule each event in $\mathcal{E}^c_2$ as a distinct task.  We then
denote the task by $\task_{M}({k_1},{k_2},i)$, where $k_1$ and $k_2$
are the sub-domains subject to an $E_2$-event update, in which two
nodes $\X^{(n)} \in V_{1}$ and $\X^{(m)} \in V_{2}$ are affected.  The
counter $i$ now denotes the total order of the $E_2$ events in
$\mathcal{E}^c_2$ as given by the model input. This implies that if
$1 \ldots n$ events exist in the window $[t_n,t_{n+1}]$, they have to
be processed by the task in this order.

Both tasks $\task_{S}$ and $\task_{M}$ are scheduled repeatedly until
the simulation reaches its end time.  Precedence dependencies between
tasks are expressed using the `$\prec$'-operator.  For example,
$\task_S(k_1,n) \prec \task_S(k_1,m)$ means that task $\task_S(k_1,n)$
must complete its execution before task $\task_S(k_1,m)$ is spawned.
Our task-based implementation contains the following dependencies:

\begin{enumerate}
\item $\task_{S}({k},n) \prec \task_{S}({k},m)$ if $n<m$, to maintain
  the causality of private updates of sub-domain $V_k$.
\item To maintain the causality of domain crossing events:
  \begin{enumerate}
  \item $\task_{M}(n) \prec \task_{M}(m)$ if $n<m$, at coarse-grained
    processing,
  \item $\task_{M}(k_a,k_b,n) \prec \task_{M}(k_c,k_d,m)$ if $n<m$ and
    $k_c \in \{k_a, k_b\}$ or $k_d \in \{k_a, k_b\}$, at fine-grained
    processing.
  \end{enumerate}
\item To maintain the causality between private sub-domain updates and
  domain-crossing events:
  \begin{enumerate}
  \item
    $\{ \task_{S}({k_1},m), \task_{S}({k_2},m), \ldots,
    \task_{S}({k_n},m)\} \prec \task_{M}(m)$
    for all sub-domains $V_1, \ldots V_n$ that will be affected by an
    $E_2$ events processed in task $\task_{M}(m)$, at coarse-grained
    processing,
  \item
    $\{ \task_{S}({k_a},n), \task_{S}({k_b},n) \} \prec
    \task_{M}({k_c},{k_b},i)$
    if $i \in [t_n,t_{n+1}]$ and $k_c \in \{k_a,k_b\}$ or
    $k_d \in \{k_a,k_b\}$, at fine-grained processing.
  \end{enumerate}
\end{enumerate}

The presented processing policies lead to a different utilization of
the task scheduler.  Firstly, the task $\task_{M}$ will be of
different size, which leads to a different synchronization behavior.
Second, rules $3(a)$ and $3(b)$ imply that a different amount of
dependencies will be created for each single task $\task_{M}$.  In the
fine-grained case, a task $\task_{M}$ is spawned when two dependencies
are met.  In the coarse-grained case, the number of dependencies per
task is set dynamically at runtime and can potentially be larger.
This can clearly have an impact on the bookkeeping overhead.

\begin{figure}
  \begin{center}
    \includegraphics[width=1\linewidth]{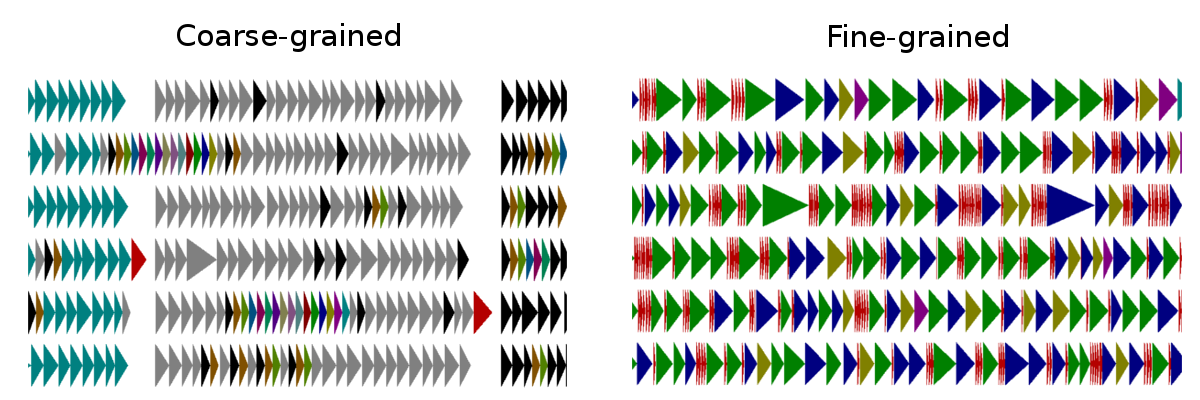}
  \end{center}
  \caption{Scheduling trace of the task-based approach; Tasks
    $\task_M$ (red color) are processing aggregated or single $E_2$
    events while tasks $\task_S$ (other colors) compute private work
    on partitioned sub-domains. As coarse-grained tasks control a
    higher number of dependencies, blocking may occur. Fine-grained
    scheduling leads to better interleaving but higher overhead
    cost. }
  \label{fig:sg_traces}
\end{figure}


\section{Computational experiments}
\label{sec:experiments}

In the following section we present results of computational
experiments of our simulator.  The following measurements were
obtained on \textit{Sandy}; a Dell Power Edge R820 computer system
equipped with four Intel Xeon E5-4650 processors and 8 cores on each
socket.  We restricted the execution to available physical cores, as
timing results on hyper-threads were strongly fluctuating. We begin
with a real-world simulation using animal movement data on national
scales, followed by a synthetic benchmark for scalability at varying
connectivity load, and we conclude with a compute-intensive parameter
estimation example.

\subsection{National scale simulation of VTEC bacteria spread}
\label{sec:vtec}

Verotoxigenic \textit{Escherichia coli} O157:H7 (VTEC O157) is a zoonotic
bacterial pathogen with the potential to cause severe disease in humans,
notably children \cite{karmali1983a, karmali1983b, riley1983}. Cattle
infected with VTEC O157 are an important reservoir for the bacteria
and they shed the bacteria in the feces without any signs of clinical
disease \cite{hancock2001}. Reducing the prevalence of infected cattle
in the population could potentially reduce the number of human
cases. However, the epidemiology of VTEC O157 in cattle is complex and
targeted interventions to control the bacteria require a thorough
understanding of the source and transmission routes
\cite{hancock2001}.

To explore the feasibility of national scale simulations to improve
the understanding of the underlying disease spread mechanisms, we have
created a model of the VTEC O157 dynamics, using the presented
framework. European Union legislation requires member states to keep
register of bovine animals including the location and the date of
birth, movements between holdings, and date of death or slaughter
\cite{Anonymous2000, Anonymous2004}. These records enable data-driven
disease spread simulations that include spatio-temporal dynamics of
the cattle population with regard to age structures, births, herd
size, slaughter, and trade patterns.

\begin{SCfigure}
  \centering
  \includegraphics[width=0.7\linewidth]{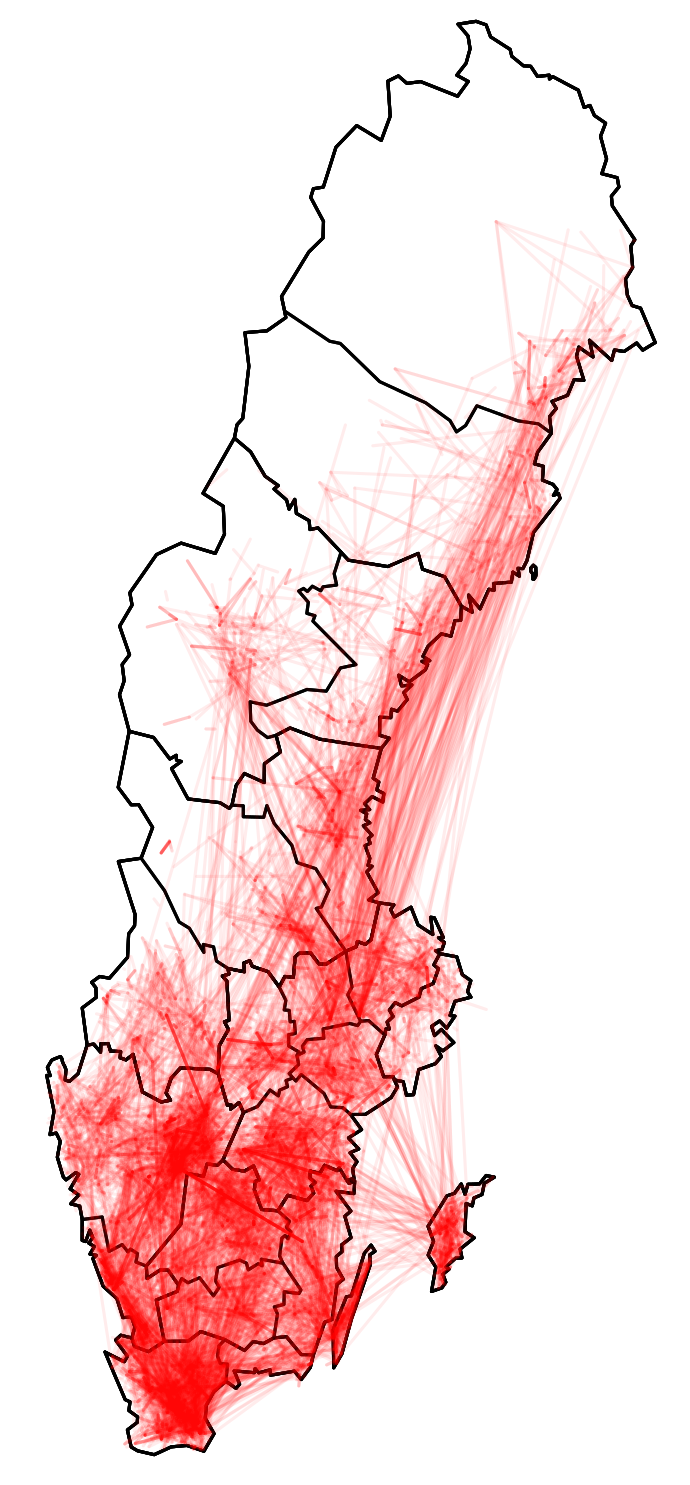}
  \caption{Visualization of cattle movements in the VTEC O157 disease
    spread simulation (\S\ref{sec:vtec}). The arcs shown are a random
    subset of the complete dataset of $\sim 10^8$ recorded events. The
    source of the data is the national cattle register at the Swedish
    Board of Agriculture.}
  \label{fig:movements}
\end{SCfigure}

The present computational experiment is based on all cattle reports to
the Swedish Board of Agriculture over the period 2005-07-01 to
2013-12-31. From these reports, three types of $E_1$ external events
(\emph{enter, exit, aging}), and a single type of $E_2$ event
(\emph{animal movements}) were condensed.  In total there were
$\sim 10^8$ external events processed during the total runtime of
$T=3106$ days. We let each integer in $0,1, \ldots, T$ represent a
synchronization window for external events, where in each window
$3707 \pm 670$ $E_1$ events and $235 \pm 104$ $E_2$ events were
processed.  A subset of the spatial network consisting of
$\Ncells=37221$ nodes is visualized in Figure \ref{fig:movements}.

Most infected cattle shed the bacteria less than 30 days before
returning to the susceptible state, but calves shed for a longer
period than adult cattle \cite{Cray1995,Davis2006}. To capture this we
let the intensity of the transitions between the states depend on the
$j$th age category,
\begin{align}
\label{eq:vtectrans}
\begin{array}{rl}
    S_j&\xrightarrow{\eta_j} I_j, \\
    I_j &\xrightarrow{\gamma_j} S_j.
  \end{array}
\end{align}
The rate for a susceptible individual on the $i$th node to become
infected per unit of time is given by
\begin{equation}
  \label{eq:rate_s_to_i}
  \eta_j = u\upsilon_{j} \varphi_{i}(t),
\end{equation}
for $i = 1,\ldots , \Ncells$ and
$j \in \{\textit{calves}, \textit{young stock}, \textit{adults}\}$. In
turn, the expected time an infected individual is in an infected state
before it returns to the susceptible state is
\begin{align}
\label{eq:recovery_rates}
  \gamma_j &= \frac{u}{\delta_j},
\end{align}
where $\delta=[28,25,22]$ and $\upsilon=[8,7,1]\times10^{-3}$ are
age-dependent constants. The factor $u$ can be understood as a
time-scale and is difficult to estimate accurately; in our experiments
it is in fact varied such that $u = 1$ closely resembles the
parameterization of the model found in \cite{siminf1}.

Finally, the continuous variable $\varphi_i$ represents the
environmental bacterial concentration that asserts an infectious
pressure on each individual at the $i$th node.  A suitable model is
given by
\begin{equation}
  \label{eq:siminfpressure}
  \frac{d \varphi_i}{dt}= \frac{\alpha \sum_{j} I_{i,j}(t)}
  {\sum_{j} S_{i,j}(t) + I_{i,j}(t)}
  - \beta(t) \varphi_i(t).
\end{equation}
Again, $i = 1, \ldots , \Ncells$ are the nodes and $S_{i,j}$ and
$I_{i,j}$ refers to the number of susceptible and infected individuals
in the $j$th age compartment at the $i$th node, respectively. The
constant $\alpha$ is the average shedding rate of bacteria to the
environment per infected individual, while $\beta$ captures the decay
and removal of bacteria. In our experiments we used the constant value
$\alpha = 1$ while $\beta(t)$ varied according to the season,
\begin{equation}
\beta(t) = \left\{
  \begin{array}{lr}
   \log(2)/14 :&  0 \le (t\bmod{365}) \le 91 \\
   \log(2)/26 :&  91 < (t\bmod{365}) \le 182 \\
   \log(2)/20 :&  182 < (t\bmod{365}) \le 273 \\
   \log(2)/12 :&  273 < (t\bmod{365}) \le 364
  \end{array}
\right.
\end{equation}

\begin{figure}
  \centering
  \begin{tabular}{cc}
    \includegraphics[width=45mm,height=45mm]{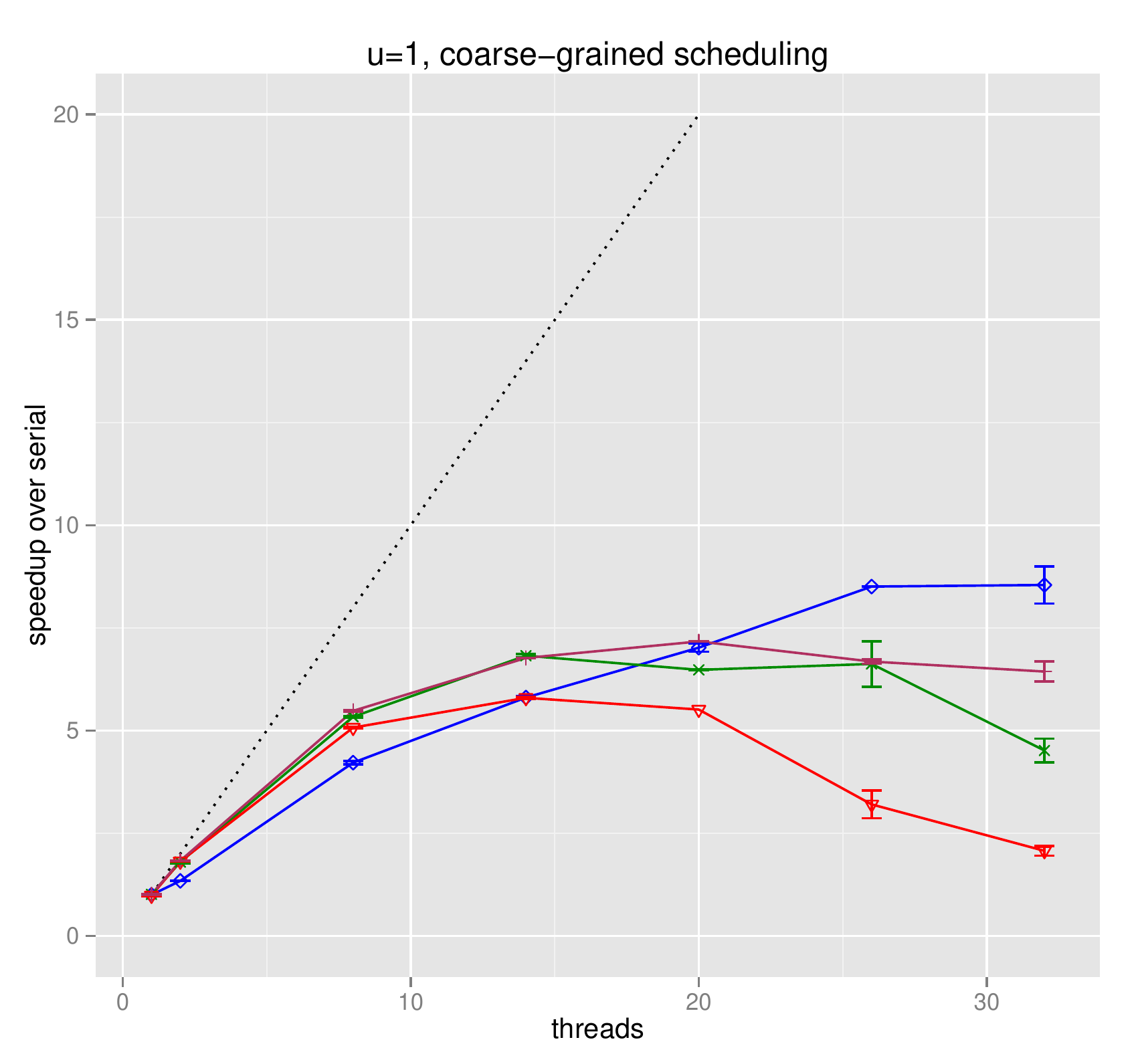} &   \includegraphics[width=54mm,height=45mm]{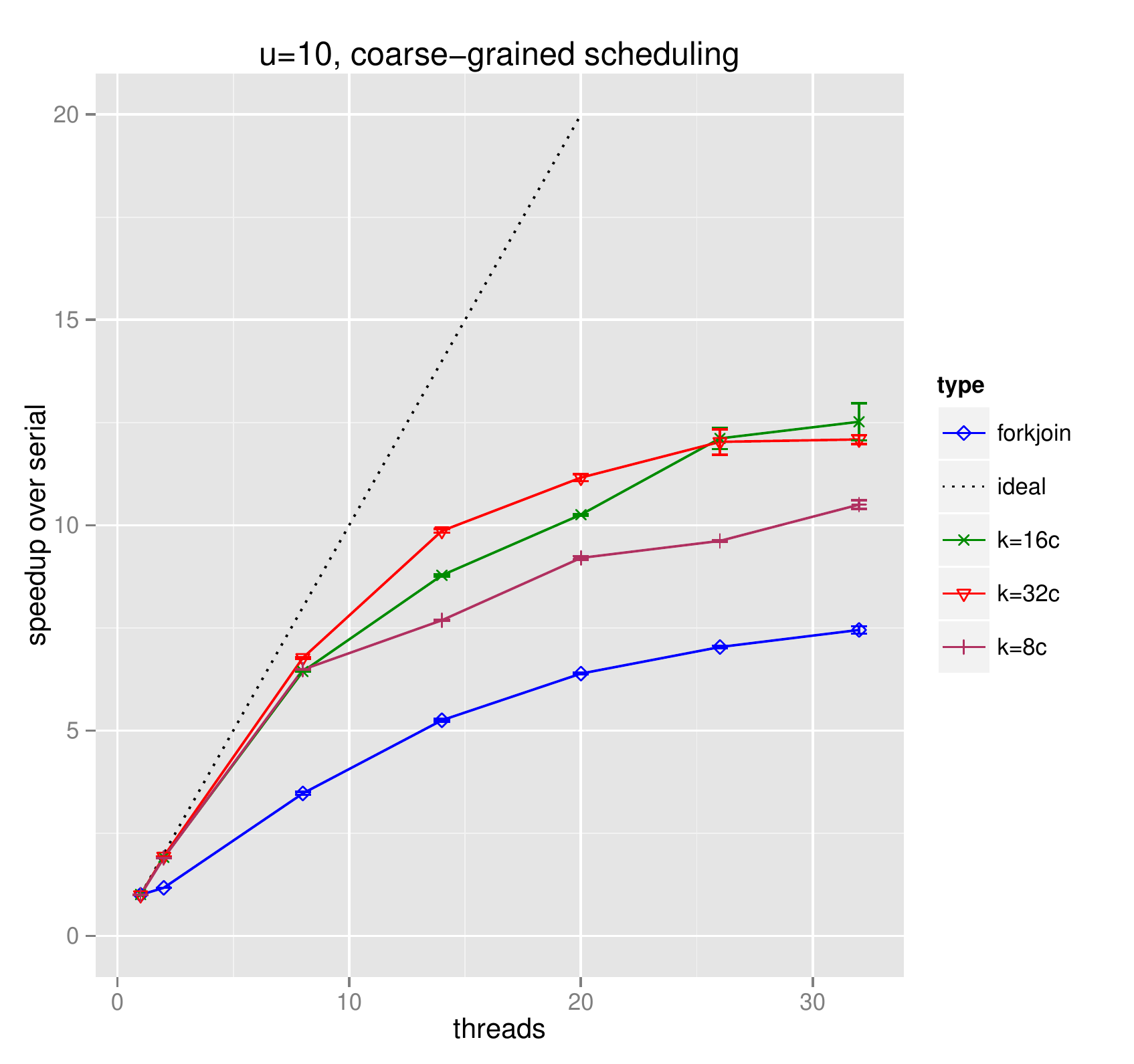} \\[6pt]
    \includegraphics[width=45mm,height=45mm]{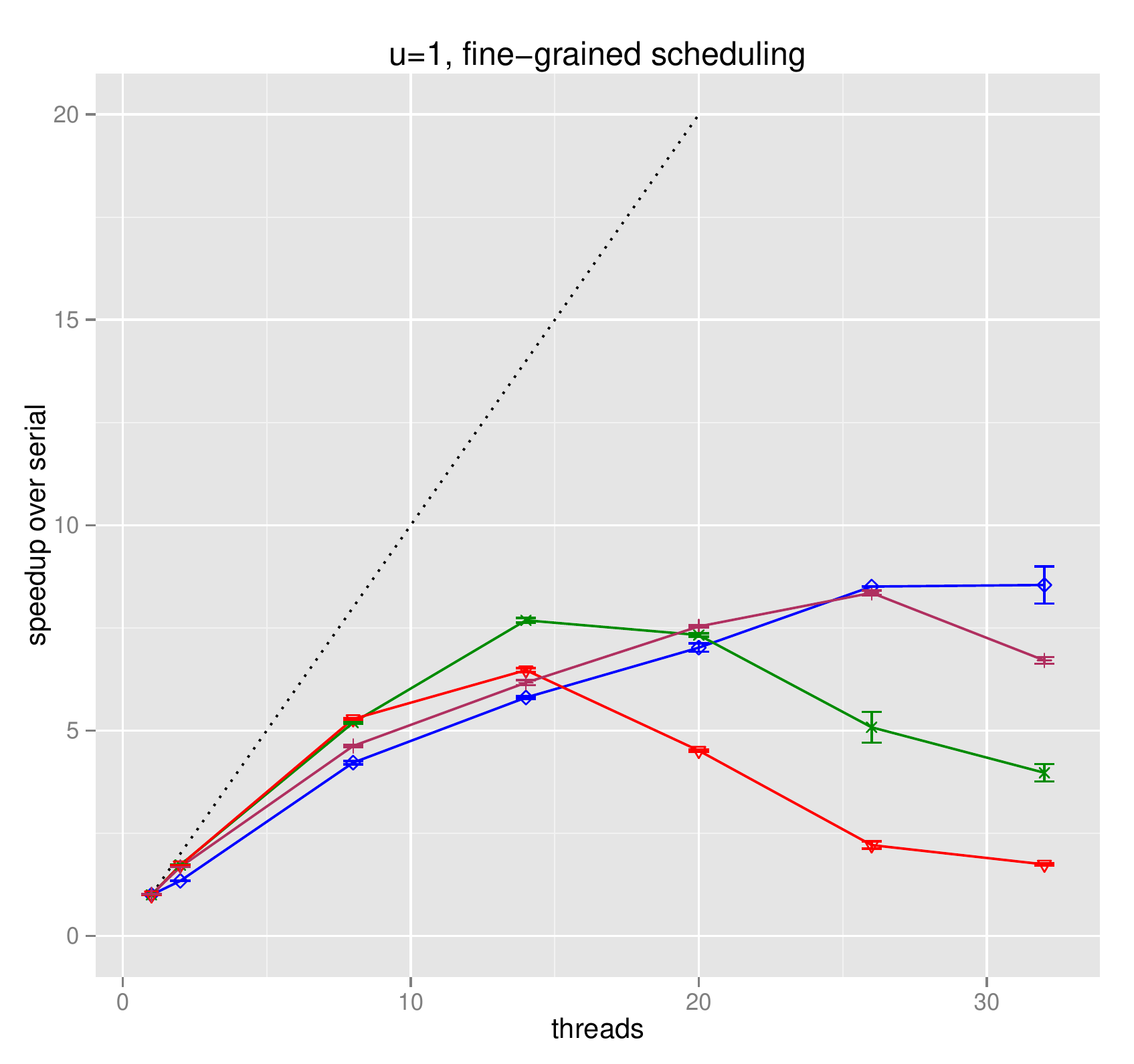} &   \includegraphics[width=54mm,height=45mm]{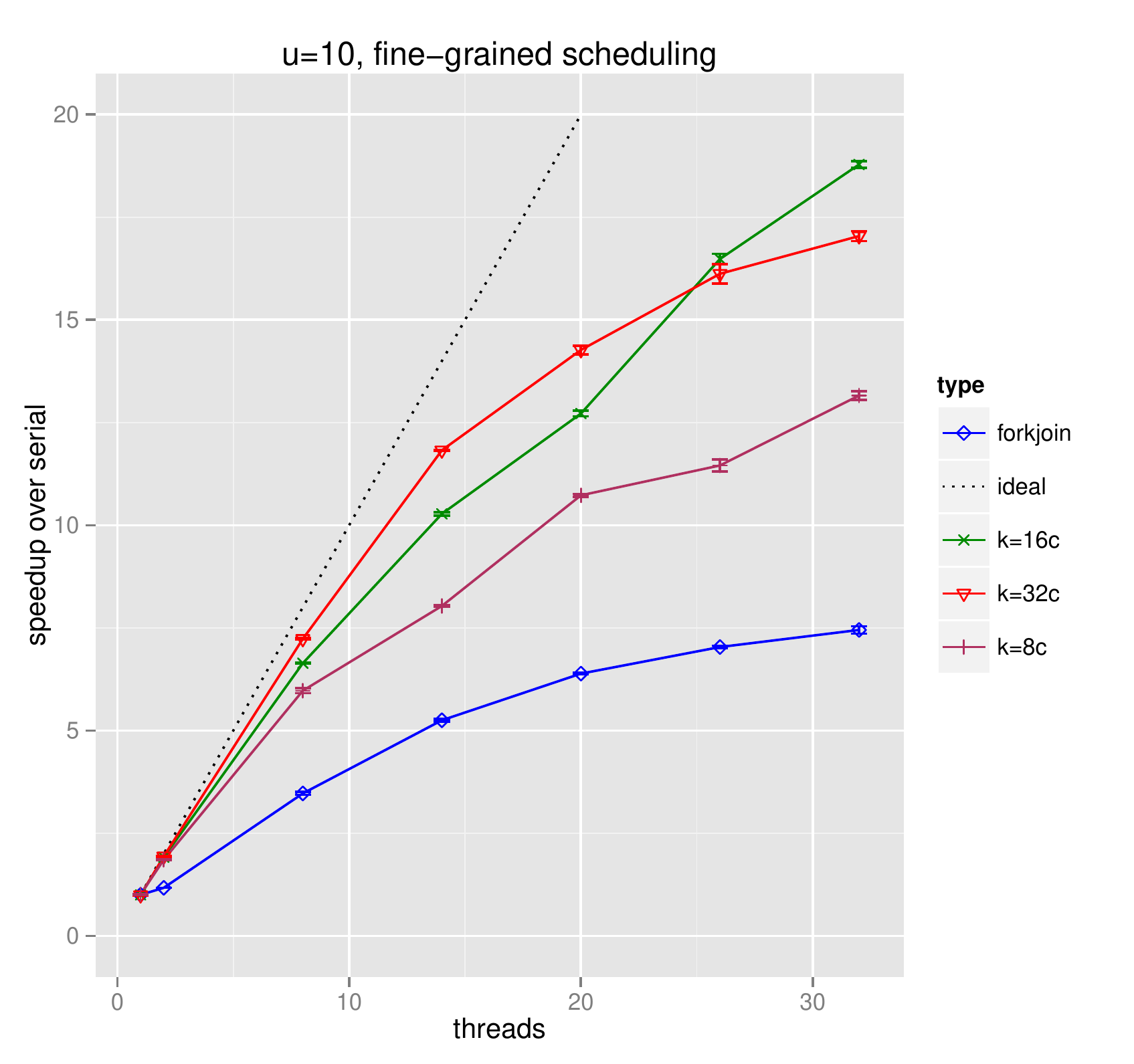} \\[6pt]
  \end{tabular}
  \caption{Performance measurements of the VTEC O157 model simulation
    on \textit{Sandy} at varying scheduling approaches, task sizes,
    and scale factor $u$. The number of tasks $k$ is chosen to be
    proportional to the number of threads $c$. In the \texttt{Open MP}
    parallelization (``forkjoin''), cross-boundary events are
    processed entirely in serial. \review{Error bars represent the
      standard error in mean (n=10).}}
  \label{fig:task_speedup}
\end{figure}

We first parallelized the simulation by spreading tasks $\task_S$ over
multiple cores using \texttt{Open MP} and serially processing the
intermediate $\mathcal{E}^c_2$ events, hereafter referred to as the
\review{fork-join} approach.  Next, we simulated the model using the
task-based approach, scheduling tasks with coarse-grained and
fine-grained policies as described in \S\ref{sec:scheduling}.  We
chose the number of sub-domains $k$ to be a multiple of the number of
threads $c$. Note that this is also the number of tasks $\task_S$
scheduled for each time window $\Delta t$.  As a higher factor $u$
creates a higher load for the tasks $\task_S$, we vary $u$ to inspect
boundary regions of the parallel performance.

\begin{figure}
  \centering
  \begin{tabular}{cc}
    \includegraphics[width=45mm,height=40mm]{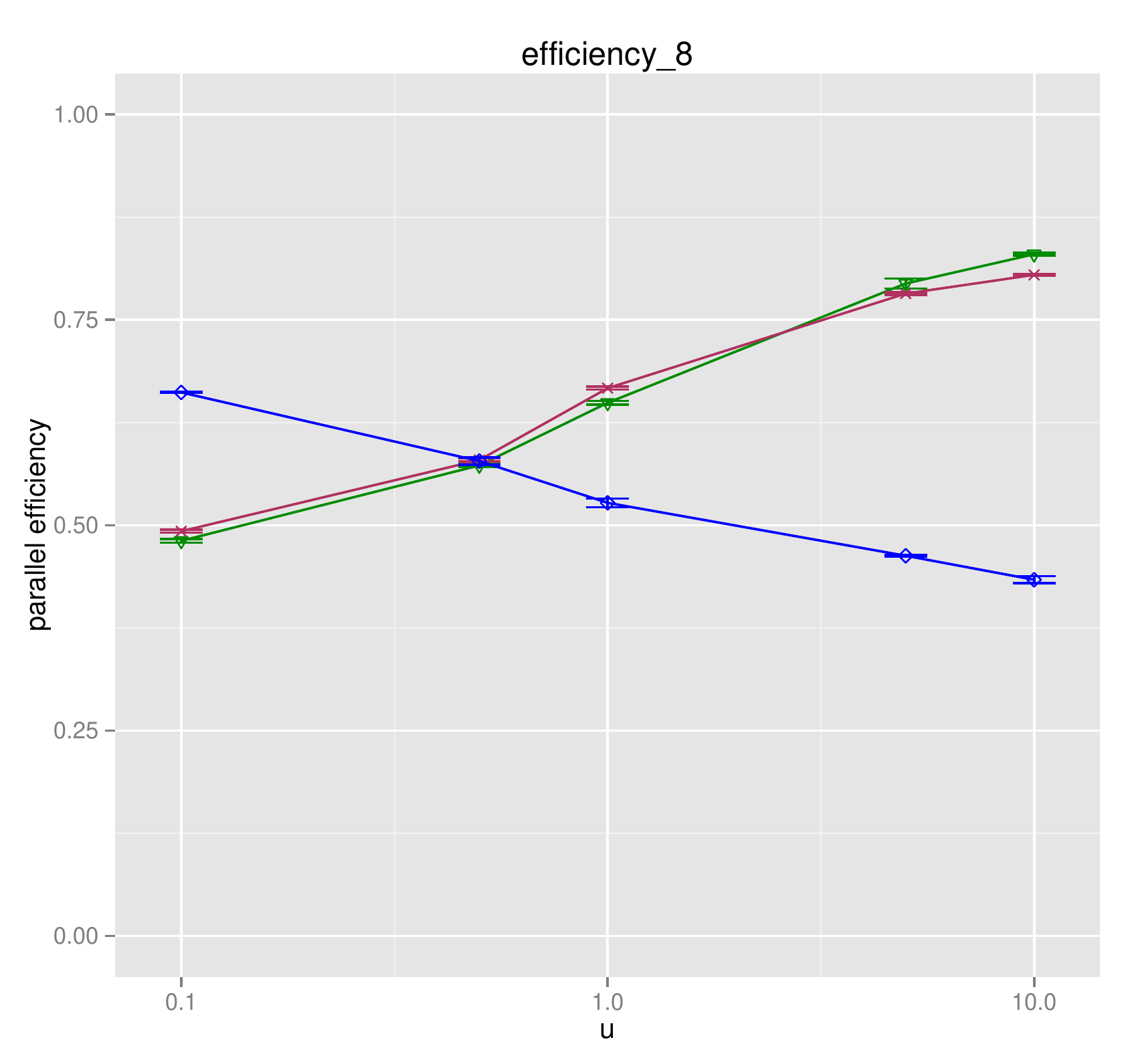} &   \includegraphics[width=45mm,height=40mm]{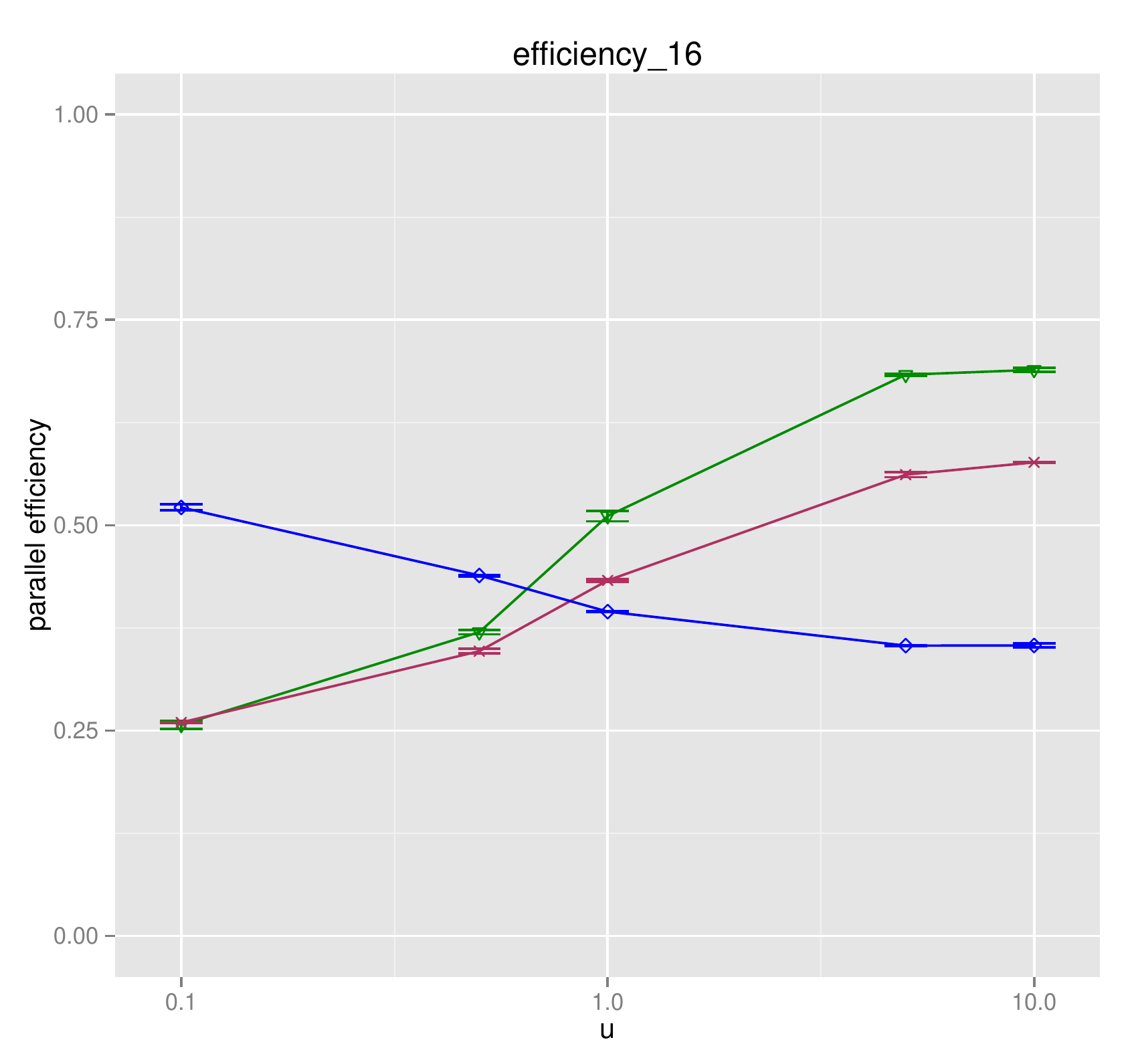} \\[6pt]
    \includegraphics[width=45mm,height=45mm]{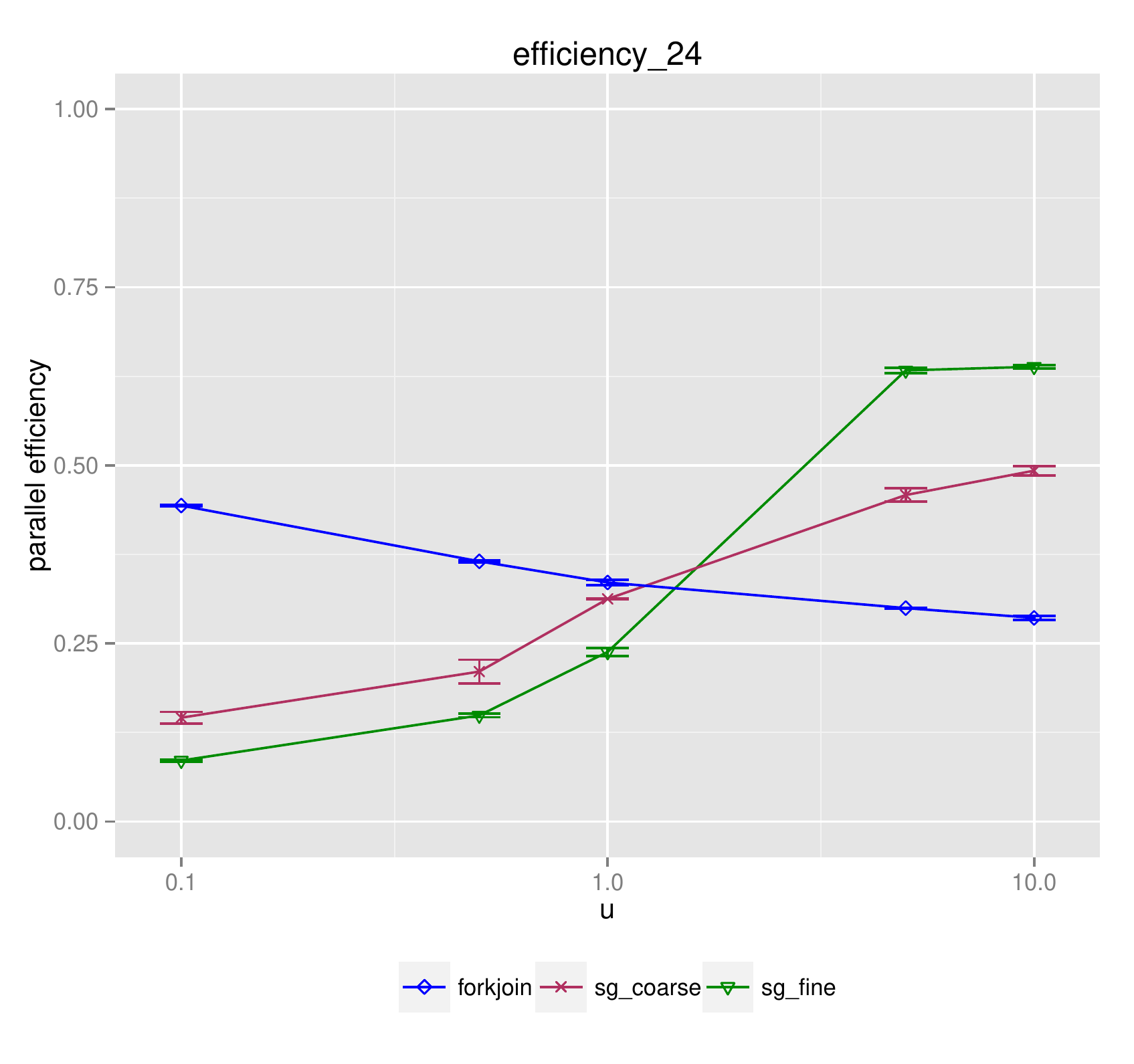} &   \includegraphics[width=45mm,height=45mm]{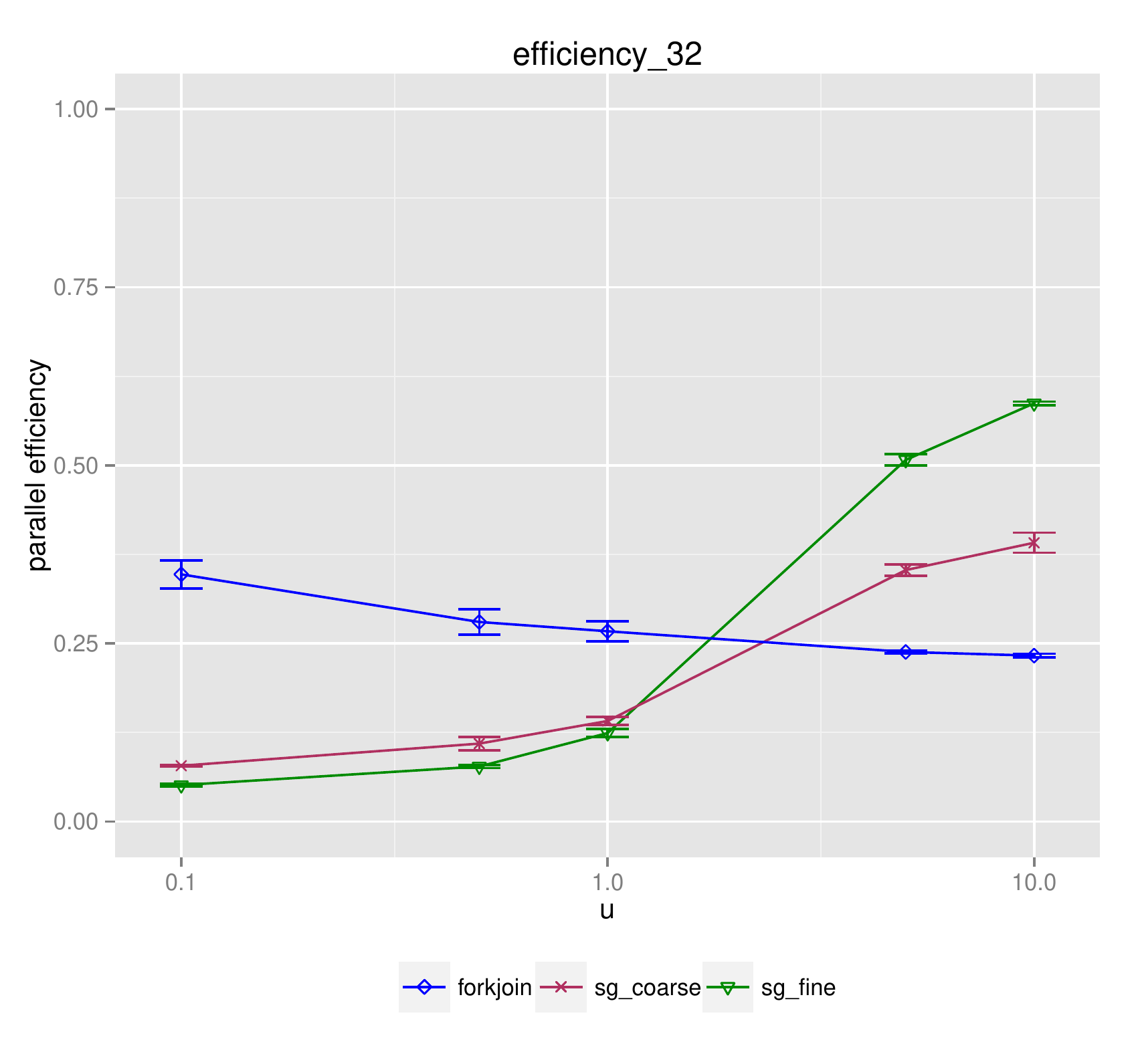}
  \end{tabular}
  \caption{Parallel efficiency of the VTEC O157 model simulation on
    \textit{Sandy} at varying factor $u$. For all task-based
    approaches the task size $k=16c$. \review{Error bars represent the
      standard error in mean (n=10).}}
  \label{fig:task_efficiency}
\end{figure}

The scaling of the different approaches is shown in
Figure~\ref{fig:task_speedup}.  For the case of $u=1$, we find that
task sizes are too small to be efficient in both task-based
approaches, and thus the \review{fork-join} approach reaches a higher
efficiency.

At $u=10$, we observe that \emph{coarse-grained} processing performs
better than the \review{fork-join} parallelization, optimally at task
sizes $k=16c$ and $k=32c$. In the case of \emph{fine-grained} task
processing, we found that the choice of $k$ has a strong impact on the
performance scaling.  While all task densities scale strongly at a
lower thread count, only the $k=16c$ density reaches a high efficiency
of $0.58$ at full thread consumption.  Thus the efficiency is more
than doubled in comparison to the efficiency of the \review{fork-join}
parallelization, which was found to be 0.23.

The dependency of the parallel efficiency on the factor $u$ is further
detailed in Figure~\ref{fig:task_efficiency}.  We observe that
scheduling overhead and small task sizes prohibit a high efficiency of
both task-based approaches if $u<1$ while the full potential of the
approaches is extractable at $u>1$ and a larger thread count.
\review{Note that the thread affinity of tasks was varied throughout
  the performed experiments in order to investigate the impact of data
  locality.}

\review{We further present a set of characteristics of the
  coarse-grained and fine-grained simulations in Table
  \ref{table:granularities} and \ref{table:timings}.  As shown in
  Table \ref{table:granularities}, the granularity of the fine-grained
  task $\task_M$ is $\sim 1/30$ of the granularity of the
  coarse-grained task $\task_M$, however the task needs to be
  scheduled about $235$ times more often throughout the simulation
  run.}

\review{On the other hand, the advantage of the fine-grained
  scheduling is emphasized by the measurements shown in Table
  \ref{table:timings}; the average waiting time to fulfill the
  dependencies for the fine-grained $\task_M$ is $44-108 \times$ lower
  than for the coarse-grained $\task_M$. This is explained by the
  larger number of dependencies associated to the coarse-grained
  $\task_M$ which is growing with the partitioning $k$.}

\begin{figure}
  \begin{center}
    \includegraphics[width=1\linewidth]{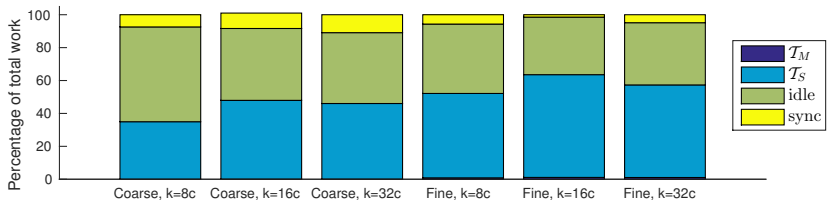}
  \end{center}
  \caption{\secreview{Percentage of total work spent on processing of
      tasks, synchronization (``sync''), and time spent waiting for
      dependencies (``idle'') for various scheduling policies when
      measured on 32 cores. Note that the relation of work to overhead
      agrees well with Figure~\ref{fig:task_speedup}.}}
  \label{fig:work}
\end{figure}

\review{The resulting execution trace is also visualized in Figure
  \ref{fig:sg_traces}, where we show that fine-grained $\task_M$ tasks
  interleave more densely with tasks $\task_S$, thus leading to lower
  idle times and higher parallel efficiency.}  \secreview{The
  percentage of total work spent on the processing of tasks, the
  synchronization of worker threads, as well as the time spent waiting
  for fulfilled dependencies are shown in Figure~\ref{fig:work} for
  the $u=10$ configuration.}

\review{For further details of
  the scheduling performance of the \texttt{SuperGlue} library in
  regards to the task sizes and the number of dependencies, we like to
  refer the reader to the benchmarks available in
  \cite{tillenius_superglue_2014}.  }

\begin{table}[htb!]
	\centering
	\begin{tabular}{ r | r r r | r r r }
	& \multicolumn{3}{c}{Task granularities ($10^3$ cycles)} & \multicolumn{3}{c}{Number of tasks}  \\
	\hline
          $k$ & $\task_S$ & $\task_M$ coarse & $\task_M$ fine & $\task_S$ & $\task_M$ coarse & $\task_M$ fine \\ \hline
          256 & 596 $\pm$ 799 & 314 $\pm$ 132 & 11 $\pm$ 5.2 & 794112 & 3102 & 731889 \\
          512 & 301 $\pm$ 461 & 328 $\pm$ 145 & 11 $\pm$ 5.2 & 1588224 & 3102 & 731889 \\
          1024 & 152 $\pm$ 294 & 327 $\pm$ 145 & 11 $\pm$ 5.2 & 3176448 & 3102 & 731889 \\ \hline
	\end{tabular}
	\caption{\review{Average task granularity $\pm$ the standard
            deviation, and the total number of tasks created during
            the simulation at a given partitioning $k$.}}
	\label{table:granularities}
\end{table}

\begin{table}[htb!]
	\centering
	\begin{tabular}{ r | r r | r r | r r r }
	& \multicolumn{4}{c}{Waiting for dependencies ($10^3$ cycles)} & \multicolumn{3}{c}{Number of dependencies}  \\
	\hline
          $k$ & $\task_S$ coarse & $\task_M$ coarse & $\task_S$ fine & $\task_M$ fine & $\task_S$ & $\task_M$ coarse & $\task_M$ fine \\ \hline
          256 & 15 $\pm$ 10 & 550 $\pm$ 200 & 12.5 $\pm$ 5 & 12.5 $\pm$ 5 & 1 & 146 $\pm$ 33 & 2 \\
          512 & 12.5 $\pm$ 5 & 850 $\pm$ 200 & 12.5 $\pm$ 5 & 12.5 $\pm$ 5 & 1 & 202 $\pm$ 58 & 2 \\
          1024 & 12.5 $\pm$ 5 & 1350 $\pm$ 500  & 11 $\pm$ 4 & 11.5 $\pm$ 4 & 1 & 248 $\pm$ 83 & 2 \\ \hline
	\end{tabular}
	\caption{\review{Maximum $\pm$ full-width-half-maximum of the
            (right-skewed) histogram of waiting times for fulfilled
            task dependencies, and the average amount $\pm$ standard
            deviation of dependencies assigned to tasks at each
            discrete time interval $[t_n,t_{n+1}]$, at a given
            partitioning $k$ on 32 computing cores.}}
	\label{table:timings}
\end{table}


\subsection{Synthetic benchmark}
\label{sec:synthetic}

The results in \S\ref{sec:vtec} indicate a delicate performance
dependency on the balance between the local events and the effective
connectivity of the network. To further investigate this a synthetic
benchmark with a \emph{fixed} load of local events was created. This
model consists of two compartments $S$ and $I$ only, both residing on
$\Ncells=1000$ nodes. The transitions are simply
\begin{align}
\begin{array}{rl}
    S&\xrightarrow{1} I, \\
    I &\xrightarrow{1} S,
  \end{array}
\end{align}
where the initial population size of each compartment $I_{i}$ and
$S_{i}$ was set to 1000. This model is considered at times
$t=[0,\Delta t, 2 \Delta t,\ldots T]$, with $\Delta t=1$ and $T=1000$,
thus generating about 2000 local events per synchronization time
window $\Delta t$ and node $i$.

The nodes were arranged into $k$ sub-domains and a total of
$\rho k(k-1)/2$ distinct $E_2$ events were generated at the end of each
time window, each connecting two randomly sampled nodes $\X^{(i)}$ and
$\X^{(j)}$ belonging to \emph{different} sub-domains. Hence $\rho = 1$
means that all sub-domains have to communicate with all other
sub-domains at each synchronization point. The number of tasks for the
coarse-grained and fine-grained approach was set to the number of
threads ($k=c$).

\begin{figure}
  \centering
  \begin{tabular}{cc}
    \includegraphics[width=0.8\linewidth]{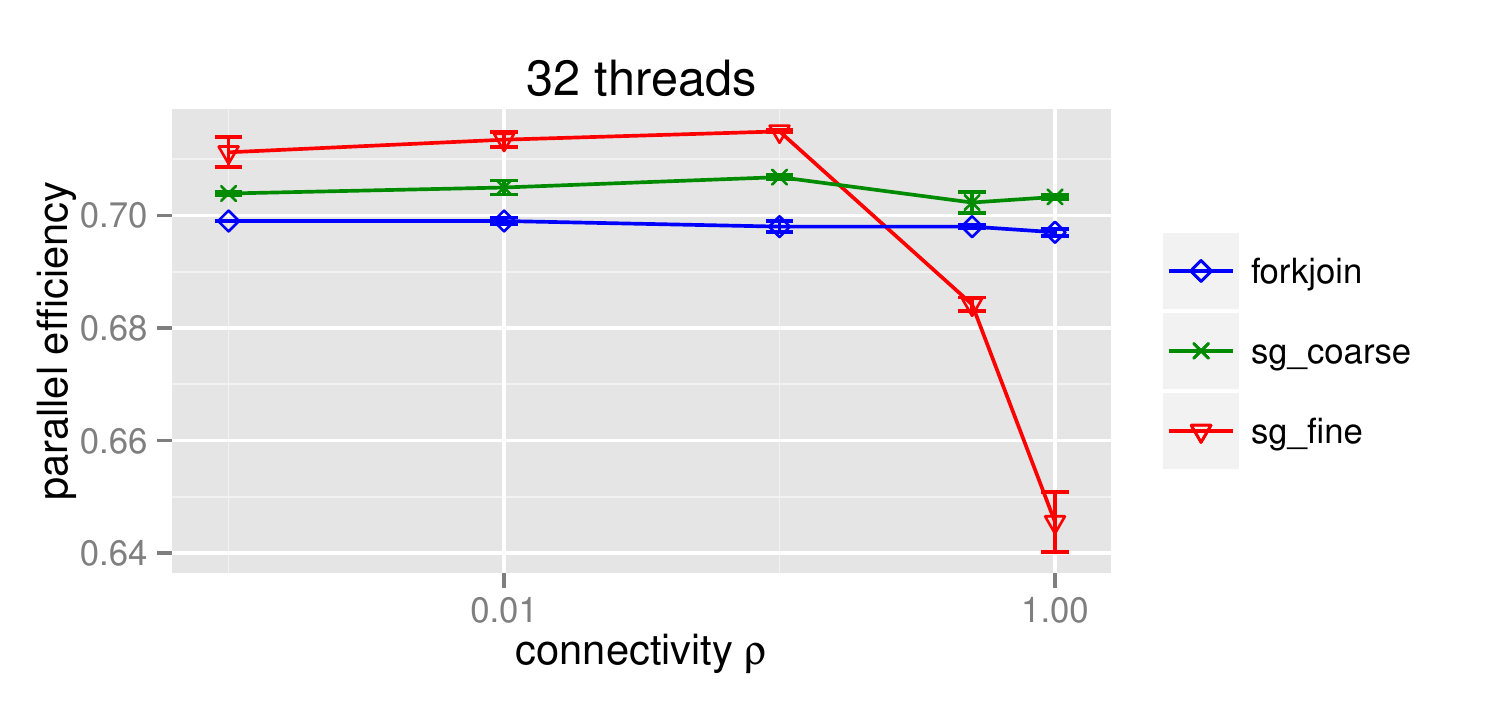}
  \end{tabular}
  \caption{Parallel efficiency for the different methods on the
    synthetic benchmark. \review{Error bars represent the standard
      error in mean (n=10).}}
  \label{fig:benchmark_efficiency}
\end{figure}

The measurements obtained on the \textit{Sandy} computer system at
full thread consumption are presented in Figure
\ref{fig:benchmark_efficiency}.  The parallel efficiency of all
methods lies at $\sim 0.7$ for $\rho \le 0.1$ and remains there even
when $\rho \to 0$ and so we deduce that the problem is memory
bound. The coarse-grained task-based implementation and the
\review{fork-join} approach scale very similarly with increasing
connectivity $\rho$. The fine-grained task-based approach attains the
highest parallel efficiency at $\rho \le 0.1$, but the performance
drops at a higher global connectivity. This phenomenon arises because
each $\task_M$-task creates dependencies on two subsequent
$\task_S$-tasks at every synchronization window, thus creating a
higher overhead and limiting asynchronous task execution.

\subsection{Feasibility of parameter estimation}
\label{sec:fitting}

A usually very compute intensive load case is the fitting of model
parameters, typically using numerical optimization of some kind. The
problem can briefly and ideally be described as follows; unknown is
the set of parameters $k^*$ and an observed time-series of data
$\X(t; \; k^*)$. The parameters $k^*$ are estimated by repeatedly
simulating a whole family of trajectories with parameters $k$, where
$k$ is modified until input data and simulations match up in some
suitable sense. The framework's feasibility of fitting an
epidemiological model is of course very important, since the modeling
process at some point or the other will involve calibration of
parameters with respect to reference data.

To demonstrate the feasibility of parameter estimation in the current
context, we use the epidemiological model introduced in
\S\ref{sec:vtec} and first identify the set of parameters
$[k_1,k_2,k_3]$ that have a high degree of \emph{observability}. A
suitable such set is
\begin{align}
  \label{eq:superparameters}
  k_j = \upsilon_{j} \delta_j,~j \in
  \{\textit{calves, young stock, adults}\}.
\end{align}
We let a reference solution be given by a single trajectory
$\X(t; \; k^*)$, with $\delta^* = [28,25,22]$,
$\upsilon = [8.8,3.2,1] \times 10^{-3}$, and with $\alpha$ and
$\beta = \beta(t)$ as in \S\ref{sec:vtec}.

To obtain a robust procedure, some kind of smoothing statistics should
be considered. We chose to aggregate counts of animals in neighboring
nodes into larger regions. To be precise, the overall domain was
divided into 21 areas (coinciding with the Swedish county codes),
after which the goodness of fit $G(k)$ was defined by
\begin{align}
\label{eq:goal}
   G(k)^2 &= \int_{0}^{T} \sum_{j} \left\| \bar{\textbf{x}}_{j}(t; \; k)-
  \sum_{l \in C(j)} \X^{(l)}(t; \; k^*) \right\|^2 \, dt, \\
  \intertext{with}
  \bar{\textbf{x}}_{j}(t; \; k) &= \frac{1}{N} \sum^N_{i=1} \bar{x}_{j}^{i}(t; \; k),
  \intertext{and where the individual sample trajectories are given by}
  \bar{x}_{j}^{i}(t; \; k)&=\sum_{l \in C(j)} \X^{(l)}(t; \; k, \; \omega_i),
\end{align}
where $N$ is the number of trajectories and
$C(j), j \in \{1,\ldots,21\}$ is the set of nodes $\X^{(l)}$ that
belong to county $j$.  To quantify the uncertainty in $G(k)$ we
compute the variance as
\begin{align}
  \label{eq:goalvariance}
  VG(k)^{2} &= \int_{0}^{T} \sum_{j} \bar{\sigma}^{2}_{j}(t; \; k) \, dt,\\
  \intertext{where}
   \label{eq:goalsigma}
  \bar{\sigma}^{2}_{j}(t; \; k) &= \frac{1}{N-1} \sum_{i}
  \| \bar{x}_{j}^{i}(t; \; k)-\bar{\textbf{x}}_{j}(t; \; k) \|^2,
\end{align}
and use $\pm 2VG(k)/\sqrt{N}$ as a measure of the uncertainty.

Next, the parameter estimation problem is approached by solving the
minimization problem
\begin{align}
\label{eq:optimum}
\hat{k} &= \arg \min_{k} G(k)^{2}.
\end{align}
In practice we make use of the \emph{Pattern Search} routine in
\cite{hooke__1961}, which conceptually resembles the Golden Section
search \cite{kiefer_sequential_1953} in its narrowing of the
search-space. The numerical optimization routine evaluates
\eqref{eq:goal} until the residual error reaches a defined
threshold. In our tests we varied the initial guess of the parameters
$k_{0}$ but found that the results did not vary substantially. In the
results below, we conveniently put $k_{0} = 1.6 k^*$.

\begin{figure}
  \begin{center}
    \includegraphics[width=1\linewidth]{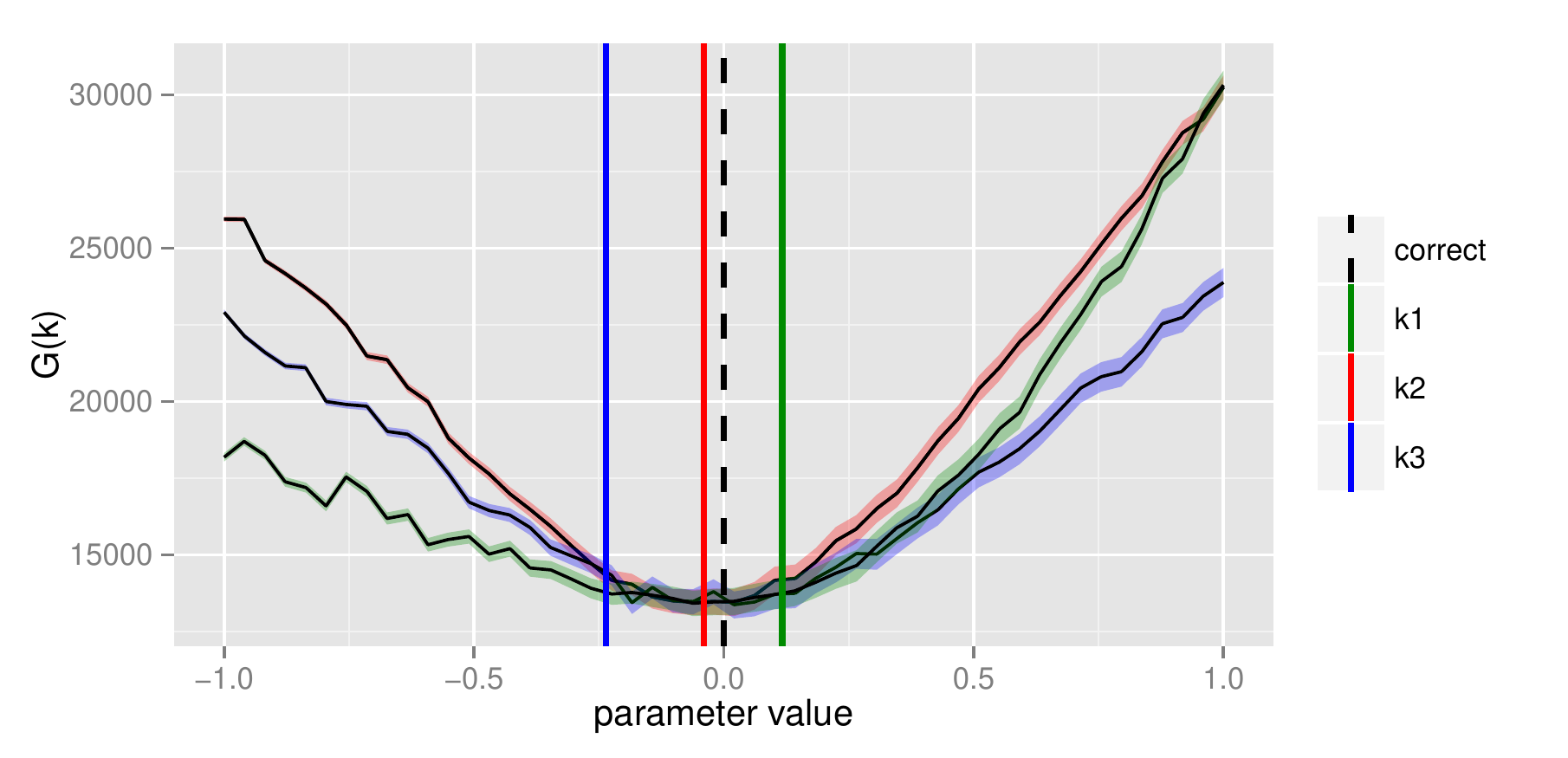}
  \end{center}
  \caption{Goal function $G(k)$ in the form of confidence intervals
    $G(k) \pm 2VG(k)$, visualized for each parameter $k_j$,
    $j=\{1,2,3\}$, when the other parameters $k_i$, $i \neq j$, are
    held at the target value $k^*_i$. Vertical lines indicate the
    target $k^*$ and the obtained estimates $k_{1 \ldots 3}$. The
    parameter ranges have been normalized for ease of comparison.}
  \label{fig:fitting_landscape}
\end{figure}

Since an increasing number of trajectories yields better estimates of
the mean and variance, we simulate using different number of
trajectories. We measure the total solver time on 12 and 32 computing
cores, respectively, and we let the total number of iterations to be
$N=20$ in all cases. The results are presented in Table
\ref{table:fitting} where the relative residual is defined as
\begin{align}
  \label{eq:residuall}
  R(k) &=\frac{| G({k}) -G(k^*) |}{| G(k^*)|}.
\end{align}
The optimization landscape of the goal function \eqref{eq:goal}, and
hence the definiteness of the setup itself, is visualized in
Figure~\ref{fig:fitting_landscape}. Due to the simple bisection search
behavior of the numerical routine, the obtained parameters $k$ are in
fact the same for all displayed cases, although the relative residuals
differ considerably.

\begin{table}[h]
	\centering
	\begin{tabular}{ r r r r }
	\hline
          Trajectories & Rel. residual & Time (c=12) & Time (c=32)  \\ \hline
          10 & 0.1738 & 46.6 min & 30.2 min \\
          20 & 0.0900 & 94.2 min & 61.5 min \\
          40 & 0.0363 & 189.3 min & 123.7 min \\ \hline
	\end{tabular}
	\caption{Solver time of the parameter estimation
          problem on 12 and 32 cores, respectively, and using a
          different number of simulated trajectories.}
	\label{table:fitting}
\end{table}

Note that the obvious approach of parallelization by computing the $N$
independent trajectories using separate threads by a sequential
algorithm is unfavorable here, for two related reasons. Firstly, each
executable needs to store a rather large state space in working
memory. Secondly, each simulation must also access the complete
database of externally scheduled events.

\section{Conclusions}
\label{sec:conclusion}

\review{Modeling and simulation are important in designing
  surveillance and control of livestock diseases and of major economic
  importance. However, various pathogens require different models to
  capture the disease dynamics and transmission routes. Moreover, an
  increasing amount of epidemiologically relevant data is becoming
  available. We have adressesed these challenges and present a
  flexible and efficient computational framework for modeling and
  simulation of disease spread on a national scale.}  The simulation
involves two parts. Firstly, the algorithm evolves stochastic dynamics
of the disease process. Secondly, a list processor incorporates
database events such as entering, exit, or movement of individuals
into the model state. The framework is highly flexible in that most
conceivable epidemiological models are either directly expressible, or
the framework may be straightforwardly extended to encompass also
non-standard models. As a concrete example it would be relatively easy
to include intervention strategies such as vaccination programs in
order to simulate the impact on the global dynamics.

We have explored different strategies to parallelize the simulator on
multi-socket architectures.  Firstly, we decomposed the spatial
information and the list of deterministic events.  We then observed
that the decomposed problem can be simulated at a high parallel
efficiency, which is limited only by the processing of cross-boundary
events. We then created three parallel implementations of the
simulator core; we used \texttt{Open MP} to only parallelize private
computations \review{in a fork-join fashion}, while cross-boundary
events were processed in serial.  Two further implementations use a
dependency-aware task scheduler to create execution traces that
interleave cross-boundary events and private computations with respect
to their dependencies.  We find that this strategy allows us to
exploit shared-memory parallelism at a higher degree than the
\review{fork-join} approach if task sizes are sufficiently
large. \review{We benchmark this approach using the \texttt{SuperGlue}
  task library, but present a set of scheduling rules defining the
  parallel simulator on general terms, thus allowing it to be
  implemented also with other dependency-aware task libraries.}

We benchmarked our simulator using a model of the spatio-temporal
spread of VTEC O157 bacteria in the Swedish cattle population. The
model contains 37221 nodes and evolves $\sim 10^8$ external events
from register data. We found that at a low private work load, the
\review{fork-join} approach performs best, mainly due to the
scheduling overhead of the task-based approaches. For higher private
work loads, the simulation benefits from task-based computing,
doubling the parallel efficiency on 32 cores in comparison to the
\review{fork-join} approach.

To further inspect the performance dependency on network properties,
we constructed a synthetic benchmark where cross-boundary events were
generated randomly. Here we found that the performance of the
\review{fork-join} approach and the coarse-grained task approach
scales well with a growing amount of cross-boundary events. Notably,
the performance of the fine-grained task processing depends more
strongly on the connectivity of boundary crossing events, thus
favoring a more fragmented network.

In a final example we used the simulator to carry out an experimental
parameter fitting within the VTEC O157 bacteria spread model. We
emphasize the high computational complexity of this task with multiple
unknown parameters to fit and the need to use several full simulation
runs to evaluate each parameter candidate. A similar load case results
when different intervention strategies are to be evaluated. For
example, even when several interventions reduce the infectious spread
globally, a policy maker could be interested in finding the most
cost-efficient strategy. With this work, we provide a powerful, highly
general and freely available software, that can contribute to a rapid
and more efficient development of realistic large-scale
epidemiological models.

Future research will encompass studies of larger inverse problems,
including more realistic data input, and more complex dynamics. Yet
another point for future study is the scalability of the task-based
approach in a distributed environment.

\section*{Acknowledgment}

We thank Martin Tillenius for providing assistance with the use of
\texttt{SuperGlue}. This work was financially supported by the Swedish
Research Council within the UPMARC Linnaeus center of Excellence
(P.~Bauer, S.~Engblom).


\bibliographystyle{unsrt}  
\bibliography{paper-2}


\clearpage

\appendix

\section{Algorithms}

\begin{algorithm}[htb!]
\caption{Sequential simulation loop}
\label{alg:main}
\begin{algorithmic}[1]

  \STATE{\textit{Initialize:} Compute all stochastic rates $\omega_{i}$
    in all nodes $\X^{(i)}, i=1,\ldots ,\Ncells$.}

  \WHILE{$t < T_{End}$}

  \FORALL{nodes i=1 \TO $\Ncells$}

  \WHILE{$t < (t_n + \Delta t)$}

  \STATE{Compute the sum $\lambda$ of all transition intensity
    functions.}

  \STATE{Sample the next stochastic event time by
    $\tau=-log(rand)/ \lambda$ using a uniformly distributed random
    variable $rand$.}

  \STATE{Determine which event happened.  Sample the next event (by
    inversion); find $n$ such that
    $\sum_{j=1}^{n-1} \omega_j(\X^{(i)}) < \lambda \, \rand \le
    \sum_{j=1}^n \omega_j(\X^{(i)})$}

  \STATE{Update the state $\X^{(i)}$ using the stoichiometric matrix
    $\Stoich$.}

  \STATE{Update $\omega_n$ using the dependency graph $G$ to
    recalculate only affected stochastic rates.}

  \ENDWHILE

  \ENDFOR

  \STATE{$t_{n+1} = t_n+ \Delta t_n$}

  \STATE{Incorporate externally defined events in lists
    $\mathcal{E}_{1,2}$. }

  \STATE{Loop over all nodes $\X^{(i)}$ and update the continuous
    state variable $Y^{(i)}$.}

  \ENDWHILE

\end{algorithmic}
\end{algorithm}


\begin{algorithm}[htb!]
\caption{Parallel simulation loop}
\label{alg:parallel}
\begin{algorithmic}[1]

  \STATE{\textit{Initialize:} Decompose the nodes $V$ into $k$
    sub-domains $V_k$. Re-arrange the external events of type $E_1$
    into private lists $\mathcal{E}^k_1$ of each sub-domain $k$, where
    all $n$ affected nodes $\X_n \in V_k$. Further divide all $E_2$
    events into the private list $\mathcal{E}^k_2$ or the list of
    domain-crossing events $\mathcal{E}^c_2$.}

  \WHILE{$t < T_{End}$}

  \FORALL{i=1 \TO $k$}

  \STATE{\% Parallel task $\task_S$;}

  \STATE{Execute line 14 of Algorithm \ref{alg:main} for all nodes in
    sub-domain $V_k$ if $n>1$.}

  \STATE{Execute lines 3-10 of Algorithm \ref{alg:main} for all nodes
    in sub-domain $V_k$ evolving time $t \in [t_n,t_n+\Delta t]$.}

  \STATE{Execute line 13 of Algorithm \ref{alg:main} for all events in
    lists $\mathcal{E}^k_1$ and $\mathcal{E}^k_2$ at time
    $t \in [t_n,t_n+\Delta t]$.}

  \STATE{\% End of parallel task $\task_S$}

  \ENDFOR

  \STATE{\% Parallel task $\task_M$;}

  \STATE{Execute line 13 of Algorithm \ref{alg:main} for all events in
    the list $\mathcal{E}^c_2$ at time
    $t \in [t_n~(t_n+\Delta t)]$.}

  \STATE{\% End of parallel task $\task_M$;}

  \STATE{$t_{n+1} = t_n+ \Delta t_n$}

  \ENDWHILE

\end{algorithmic}
\end{algorithm}

\end{document}